\documentclass[12pt,aip,pop,reprint]{revtex4-2}

\usepackage{graphicx}
\usepackage{hyperref}
\hypersetup{colorlinks=true,linkcolor=blue,citecolor=blue}
\usepackage{siunitx}
\usepackage{mathtools}

\begin{document}

\title{Ion-acoustic feature of collective Thomson scattering in non-equilibrium two-stream plasmas}

\author{K. Sakai}
\email[]{kentaro.sakai@eie.eng.osaka-u.ac.jp}
\affiliation{Graduate School of Engineering, Osaka University, 2-1 Yamadaoka, Suita, Osaka 565-0871, Japan.}

\author{T. Nishimoto}
\affiliation{Graduate School of Engineering, Osaka University, 2-1 Yamadaoka, Suita, Osaka 565-0871, Japan.}

\author{S. Isayama}
\affiliation{Department of Advanced Environmental Science and Engineering, Kyushu University, 6-1 Kasuga-Koen, Kasuga, Fukuoka 816-8580, Japan.}

\author{S. Matsukiyo}
\affiliation{Department of Advanced Environmental Science and Engineering, Kyushu University, 6-1 Kasuga-Koen, Kasuga, Fukuoka 816-8580, Japan.}

\author{Y. Kuramitsu}
\email[]{kuramitsu@eei.eng.osaka-u.ac.jp}
\affiliation{Graduate School of Engineering, Osaka University, 2-1 Yamadaoka, Suita, Osaka 565-0871, Japan.}

\date{\today}

\begin{abstract}
We theoretically and numerically investigate the ion-acoustic features of collective Thomson scattering (CTS) in two-stream plasmas. When the electron distribution functions of two (stationary and moving) components overlap with each other at the phase velocities corresponding to the two resonant peaks of the ion-acoustic feature, the theoretical spectrum shows asymmetry because the rate of electron Landau damping is different for the two peaks. The results of numerical simulations agree well with the theoretical spectra. We also demonstrate the effect of a two-stream-type instability in the ion-acoustic feature. The simulated spectrum in the presence of the instability shows an asymmetry with the opposite trend to the overlapped case, which results from the temporal change of the electron distribution function caused by the instability. Our results show that two-stream plasmas have significant effects on CTS spectra and that the waves resulting from instabilities can be observed in the ion-acoustic feature. 

\end{abstract}

\maketitle

\section{Introduction}

Collisionless shocks are known to accelerate charged particles to nonthermal energies. The formation of collisionless shocks is mediated by electromagnetic fields due to collective plasma effects rather than Coulomb collisions. Since the width of a shock transition layer is smaller than the Coulomb mean free path, multiple plasmas with different origins can coexist, especially in the upstream regions of collisionless shocks. \cite{burgess12ssr} Some of the upstream particles are reflected at the shock front, and the upstream and reflected plasmas form a two-stream state. The free energy of the two-stream state excites plasma waves, and particles can be accelerated by wave-particle interactions. \cite{burgess12ssr,matsukiyo06jgr} In some cases, the excited waves self-organize into global structures in collisionless shocks. \cite{umeda14pop,lowe03angeo} However, it is difficult to observe the multiscale nature of collisionless shocks in space and astrophysical plasmas. This is because it is challenging to obtain images of shocks with spacecraft in space plasmas and there are no in-situ observations of distribution functions in astrophysical plasmas. Recently, collisionless shocks have been studied with laboratory experiments that have the potential to measure both global and local information simultaneously.\cite{takabe21hpl,kuramitsu12ppcf}

In laboratory experiments, collective Thomson scattering (CTS) is widely used to measure local plasma parameters and distribution functions by observing the dispersion properties and amplitudes of waves.\cite{froula11} CTS is a parametric resonance that occurs among incident electromagnetic, scattered electromagnetic, and plasma waves. There are two kinds of spectra in unmagnetized plasmas --- resonances with Langmuir waves create electron features, and those with ion-acoustic waves create ion-acoustic features. Both of these have two peaks, associated with waves propagating toward and against the observation direction. In an electron feature, the wavelength difference between peaks, the widths of the peaks, and the Doppler shift correspond to the electron density, temperature, and velocity, respectively, when the distribution function is Maxwellian. In an ion-acoustic feature, the wavelength difference between peaks, the widths of the peaks, the Doppler shift, and the asymmetry in the peaks correspond to the sound velocity, ion temperature, ion velocity, and velocity difference between the electrons and ions, respectively. It is known that the dynamic structure factor represents the CTS spectrum in quasi-equilibrium plasmas. \cite{froula11} The CTS spectra of non-equilibrium plasmas are still a subject of investigation \cite{milder21prl,milder21pop,turnbull20nphys,henchen18prl,sakai20pop,matsukiyo16jpcs,saito00angeo} and the analysis of CTS spectra normally assumes the distribution function to be Maxwellian. However, non-equilibrium plasmas with distributions that are far from Maxwellian are ubiquitous in high-energy phenomena, such as collisionless shocks. In collisionless shock experiments, CTS can be used to observe the structure of shocks and sometimes two components of the ion feature, which correspond to the two-stream state. \cite{yamazaki22pre,schaeffer19prl,swadling20prl,ross12pop,morita13pop,rinderknecht18prl} Although non-equilibrium plasmas are essential in shock formation and particle acceleration, most analyses assume a Maxwellian distribution function. Therefore, in this work, we investigate CTS for non-equilibrium plasmas and the resulting waves associated with collisionless shocks. 

We have previously investigated the electron features of CTS in two-plasma states. \cite{sakai20pop} When the relative drift of two plasmas is larger than the electron thermal velocity, the CTS spectrum is not explained by the dynamic structure factor and shows a large asymmetry reflecting the directional wave generated by the two-stream-type instability. In this paper, we focus on the ion-acoustic features of CTS in two plasmas. We theoretically calculate the CTS spectra in the two-plasma state in Sec. \ref{sec_theory}. When two electron distribution functions overlap with each other, spectral asymmetry arises due to the different rates of electron Landau damping for the two peaks. We perform numerical simulations to calculate the CTS spectra in Sec. \ref{sec_sim}. In the presence of the electron two-stream-type instability, we observe a new spectrum corresponding to the excited wave and the spectral asymmetry of the ion-acoustic feature due to the change in the electron distribution function. We provide a discussion and summary of our research in Sec. \ref{sec_discussion}.

\section{Dynamic structure factor in a two-plasma state}
\label{sec_theory}

We theoretically calculate the CTS spectrum in a two-plasma state. We consider the electron and ion distribution functions to be the sum of two Maxwell distributions: 
\begin{equation}
f_{e,i} (v) = \sum_j \frac{n_{e,i}^{j}}{n_{e,i}} f_{e,i}^{j}(v), 
\label{eq_distribution}
\end{equation}
where $f_{e,i}^{j}(v)$ and $n_{e,i}^{j}$ are the distribution function and number density of the $j$-th plasma species, respectively. The total number density is given by $n_{e,i} = \sum_{j} n_{e,i}^{j}$. The individual distribution functions are written as $f_{e,i}^{j}(v) = \exp [-(v-v_{dj})^{2}/v_{te,i}^2]/(\pi^{1/2} v_{te,i})$, where $v_{dj}$ is the $j$-th drift velocity. We assume there are no net currents, so $J=\sum_j [(q_{e} n_{e}^{j} + q_{i} n_{i}^{j})v_{dj}]=0$, where $q_{e,i}$ is the charge of an electron or ion. The thermal velocity is given by $v_{te,i} = (2 k_{B} T_{e,i}/m_{e,i})^{1/2}$, where $k_{B}$, $T_{e,i}$, and $m_{e,i}$ are the Boltzmann constant, temperature, and mass, respectively. In this paper, we assume the temperatures of the two components (electrons and ions) are the same for simplicity. We set the ion-to-electron mass ratio to 100. The reduced mass is employed to compare the dynamic structure factor with the simulated spectrum in Sec. \ref{sec_sim}. The dynamic structure factor with a realistic mass ratio is presented in Appendix \ref{sec_massratio}. 
The dynamic structure factor that describes the spectral shape of Thomson scattering near the equilibrium plasmas is expressed as \cite{froula11}
\begin{equation}
S(\mathbf{k},\omega) = \frac{2\pi}{k} \left[\left|1-\frac{\chi_{e}}{\varepsilon}\right|^2 f_{e}\left(\frac{\omega}{k}\right) + Z\left|\frac{\chi_{e}}{\varepsilon}\right|^2 f_{i}\left(\frac{\omega}{k}\right)\right]. 
\label{eq_form}
\end{equation}
The first and second terms on the right-hand side in Eq. \eqref{eq_form} are associated with the non-collective scattering or Langmuir (electron plasma) waves and ion-acoustic waves, respectively. The susceptibility $\chi_{e,i}$ is given by 
\begin{equation}
\begin{split}
\chi_{e,i} = & \frac{4\pi q_{e,i}^2 n_{e,i}}{m_{e,i} k^2} \int_{-\infty}^{\infty} \frac{\frac{\partial f_{e,i} (v)}{\partial v}}{\frac{\omega}{k}-v}dv \\
= & -\frac{4\pi q_{e,i}^2}{m_{e,i}k^2}\sum_{j}\left[\frac{n_{e,i}^{j}}{v_{te,i}^2}Z' \left(\frac{\frac{\omega}{k}-v_{dj}}{v_{te,i}}\right)\right], 
\label{eq_susceptibility}
\end{split}
\end{equation}
where $Z'(\zeta)$ is the derivative of the plasma dispersion function. The wavenumber and frequency of the observed density fluctuation are given by $\mathbf{k}=\mathbf{k_S}-\mathbf{k_I}$ and $\omega=\omega_S-\omega_I$, where $\mathbf{k_S}$, $\mathbf{k_I}$, $\omega_S$, and $\omega_I$ are the scattered wavenumber, incident wavenumber, scattered frequency, and incident frequency, respectively. The dielectric function satisfies $\varepsilon=1+\chi_e+\chi_i$. The peaks of the CTS spectrum are mainly determined by $\varepsilon\sim 0$, which represents the resonant condition of parametric resonance. The dispersion relations of the Doppler-shifted ion-acoustic and electromagnetic waves are given by $(\omega-\mathbf{k}\cdot \mathbf{v_d})^2 = c_S^2 k^2$ and $\omega^2=\omega_{p}^2+c^2k^2$, where $\mathbf{v_d}$, $c_S$, and $\omega_p$ are the flow velocity, sound velocity, and plasma frequency, respectively. Considering the dispersion relations and the resonant condition, we find the peak wavenumbers of the CTS spectra $k_{S\pm}$:
\begin{equation}
\begin{split}
k_{S\pm}^2 \simeq & \left[\frac{c^2 + (v_d \pm c_S)^2}{c^2 - (v_d \pm c_S)^2} k_I \sin{\frac{\theta}{2}} + \frac{2(v_d \pm c_S)}{c^2 - (v_d \pm c_S)^2} \omega_{I}\right]^2 \\
& + k_I^2 \cos^2{\frac{\theta}{2}},
\label{eq_peak}
\end{split}
\end{equation}
where $\theta$ is the angle between the incident and scattered wavenumbers. The peaks are determined not only by $\varepsilon\sim 0$ but also by the electron susceptibility in the numerator in Eq. \eqref{eq_form}. Substituting the dielectric function into the absolute value term in Eq. \eqref{eq_form}, we get
\begin{equation}
\left|\frac{\chi_{e}}{\varepsilon}\right|^{2} = \left|\frac{1}{1+x}\right|^{2} = \frac{1}{(1+\mathrm{Re} [x])^2 + (\mathrm{Im} [x])^2}, 
\label{eq_parametric}
\end{equation}
where $x=(1+\chi_i)/\chi_e$. The peak is determined to be at the wavenumber with the smallest denominator in Eq. \eqref{eq_parametric} where $\mathrm{Re} [x]\sim-1$ and $\mathrm{Im} [x]\sim 0$ are satisfied. 

\begin{figure*}
    \centering
    \includegraphics[clip,width=\hsize]{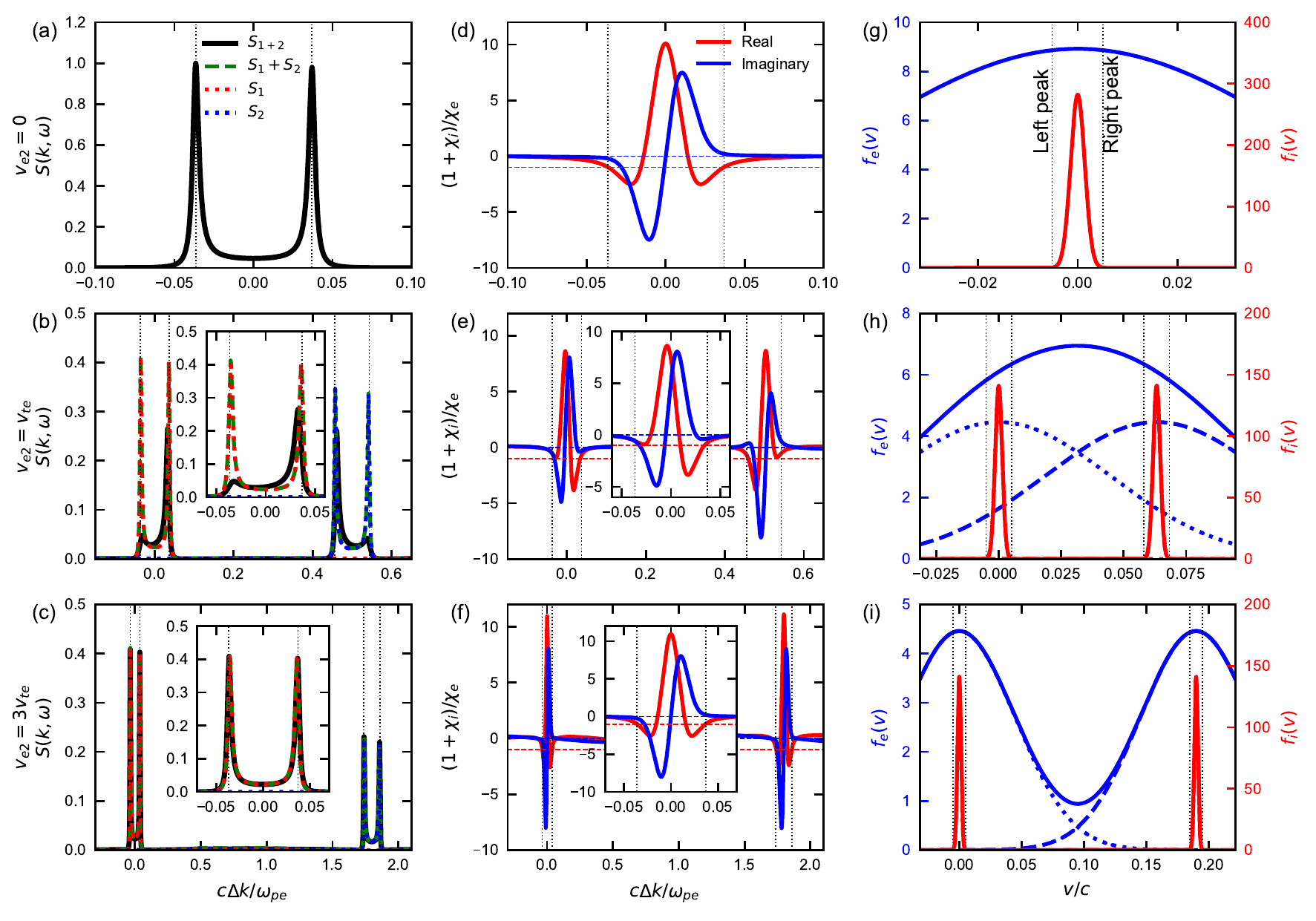}
    \caption{The dynamic structure factor in the two-plasma state. (a)--(c) CTS spectra, (d)--(f) $(1+\chi_i)/\chi_e$, (g)--(i) Distribution functions.}
    \label{fig_1}
\end{figure*}

We plot Eqs. \eqref{eq_distribution}, \eqref{eq_form}, and \eqref{eq_parametric} in Fig. \ref{fig_1} with the parameters $T_{e}/(m_{e}c^{2})=2\times 10^{-3}$, $T_{e1}=T_{e2}=10T_{i1}=10T_{i2}$, $n_{e,i}^{1}/n_{e,i}^{2}=1$, $Z=1$, $v_{d1}=0$, $ck_{I}/\omega_{pe}=5$, and $\theta=\SI{90}{\degree}$. We compare the spectra obtained from $f=f^{1}+f^{2}$ ($S_{1+2}$) and the sum of the spectra obtained independently from $f_{1}$ ($S_{1}$) and $f_2$ ($S_{2}$) by changing the drift velocity $v_{d2}/v_{te}= 0$, 1, and 3. Figures \ref{fig_1}(a)--\ref{fig_1}(c) show the scattered spectrum in terms of the wavenumber shift, $\Delta k = k_S - k_I$, normalized to $\omega_{pe}/c$, where $\omega_{pe}$ is the electron plasma frequency. The black, green dashed, red dotted, and blue dotted curves in Figs. \ref{fig_1}(a)--\ref{fig_1}(c) show the CTS spectra of $S_{1+2}$, $S_{1}+S_{2}$, $S_{1}$, and $S_{2}$, respectively. The red and blue curves in Figs. \ref{fig_1}(d)--\ref{fig_1}(f) are the real and imaginary parts of $(1+\chi_i)/\chi_e$ in terms of the wavenumber shift, respectively. The resonant condition of $\mathrm{Re} [(1+\chi_i)/\chi_e] \sim -1$ is shown by the horizontal red dashed line, and that of $\mathrm{Im} [(1+\chi_i)/\chi_e] \sim 0$ is shown as the horizontal blue dashed line. We plot Eq. \eqref{eq_peak} as the vertical dotted lines in Figs. \ref{fig_1}(a)--\ref{fig_1}(f) as a criterion of the peaks determined by Eq. \eqref{eq_parametric}. The insets in Figs. \ref{fig_1}(b), \ref{fig_1}(c), \ref{fig_1}(e), and \ref{fig_1}(f) are enlarged versions of the left spectrum for ease of view. Figures \ref{fig_1}(g)--\ref{fig_1}(i) represent the electron (blue) and ion (red) distribution functions. The solid, dotted, and dashed curves show $f=f^{1}+f^{2}$, $f^{1}$, and $f^{2}$, respectively. The vertical dotted lines indicate the phase velocity corresponding to the peaks in Eq. \eqref{eq_peak}. 

Figures \ref{fig_1}(a), \ref{fig_1}(d), and \ref{fig_1}(g) show the results when $v_{d2}=0$, i.e., a single Maxwellian distribution. There are two symmetric peaks of $S_{1+2}$ because of the isotropic ion-acoustic fluctuations in the single Maxwellian distribution. 
Figures \ref{fig_1}(b), \ref{fig_1}(e), and \ref{fig_1}(h) show the results when $v_{d2}/v_{te}=1$. The ion distribution functions do not overlap with each other because the ion thermal velocity is $\sim 32$ times lower than the electron thermal velocity. However, the electron distribution functions overlap and form the higher-temperature ($\sim 1.6 T_e$) and moving ($\sim v_d/2$) single Maxwellian-like distribution. In the corresponding spectrum, there are two ion-acoustic features; one comes from the stationary ($j=1$) plasma that has two peaks around $c\Delta k/\omega_{pe} \sim 0$, and the other from the moving ($j=2$) plasma that has two peaks around $c\Delta k/\omega_{pe} \sim 0.5$ (blue-shifted). One can see the right peak of the stationary plasma and the left peak of the moving plasma in the $S_{1+2}$ spectrum are stronger than the other peaks, while the peaks in the $S_{1}+S_{2}$ spectrum are symmetric. The plot of $(1+\chi_i)/\chi_e$ shows the stronger peaks are closer to the resonant condition than the weaker ones. As the ion distribution functions do not overlap, the asymmetry comes from the electron distribution function. The spectral asymmetry arises when the rates of the electron Landau damping for the two ion-acoustic peaks are different, i.e., when the slopes of the electron distribution function are different for the resonant phase velocities. Assuming a Maxwellian distribution, the asymmetry can be interpreted as the velocity difference between the electrons and ions. \cite{froula11} The derivative of the electron distribution function is different for each peak, and the derivatives of the stronger peaks are smaller than those of the weaker peaks. The larger the derivative of the electron distribution, the stronger the damping. Therefore, the spectral asymmetry originates from the different damping rates for the two peaks considering the total electron distribution function. When comparing the two ion-acoustic features of the stationary and moving plasmas, the intensity of the spectrum from the stationary plasma is larger than that of the spectrum from the moving plasma because the $\mathbf{k}$-vector is larger and $\alpha$ is smaller for the spectrum from the moving plasma. 
Figures \ref{fig_1}(c), \ref{fig_1}(f), and \ref{fig_1}(i) show the results when $v_{d2}/v_{te}=3$. The electron and ion distribution functions both do not overlap at the phase velocities corresponding to the peaks. The ion-acoustic feature from the moving ($j=2$) plasma peaks around $c\Delta k/\omega_{pe} \sim 1.8$. The spectral shape of $S_{1+2}$ is similar to that of $S_{1}+S_{2}$ because the shapes of the distribution functions, i.e., the rate of Landau damping, are almost the same as those for a single Maxwellian distribution at the peak phase velocity. 
\begin{figure}
    \centering
    \includegraphics[clip,width=\hsize]{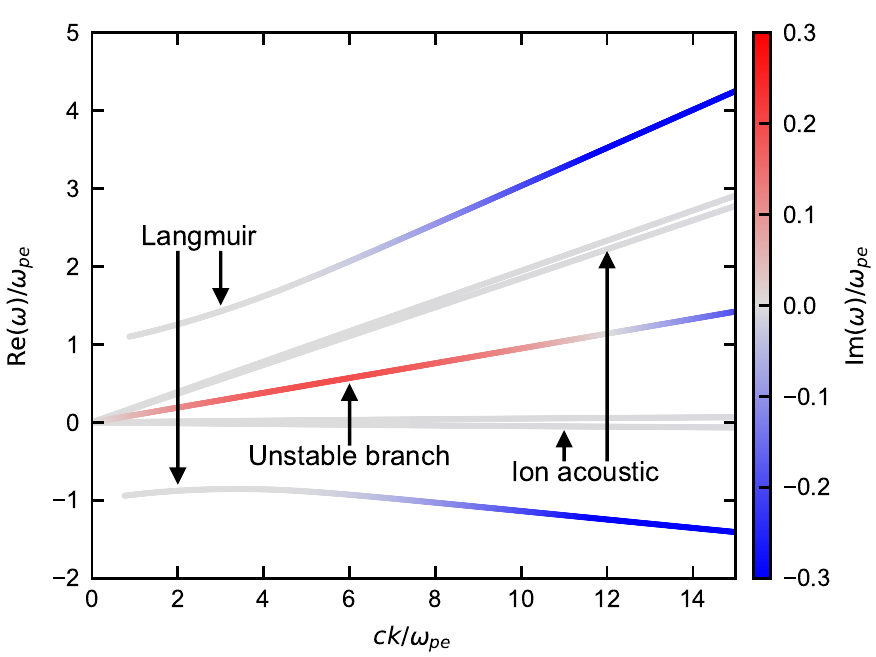}
    \caption{The dispersion relation for the distribution function in Fig \ref{fig_1}(i).}
    \label{fig_2}
\end{figure}
However, in such a case ($v_{d2}\gtrsim v_{te}$), the effect of the two-stream-type instability should be considered with the temporal variation of the electron distribution function. \cite{matsukiyo06jgr,sakai20pop} We plot the dispersion relation for the case of $v_{d2}\gtrsim v_{te}$ [Fig. \ref{fig_1}(i)] in Fig. \ref{fig_2}. The color of each branch represents the growth rate. The growth rate of the two-stream-type instability is positive at $\omega/k\sim v_{d2}/2$ and is maximized around $(ck/\omega_{pe}, \omega/\omega_{pe}) \sim (5.6, 0.53)$. This results in the wave excitation and temporal variation of the distribution function. Numerical simulations are required to treat the effect of an unstable electron distribution function on the ion-acoustic feature. 

\section{Numerical simulations of a scattered spectrum}
\label{sec_sim}

As the plasma state varies over time in non-equilibrium plasmas, we carried out numerical simulations to obtain the CTS spectra including the effects of the time-dependent distribution functions and instabilities. We employed a numerical method in which the wave equation of scattered light is solved to obtain the CTS spectra. \cite{sakai20pop,matsukiyo16jpcs} Since our focus was the ion-acoustic feature of Thomson scattering, we included both the effects of the electron and ion dynamics on the density fluctuation and solved the following wave-coupling equation:
\begin{equation}
\left(-\nabla^2 + \frac{1}{c^2} \frac{\partial^2}{\partial t^2} + \frac{\omega_{p}^2}{c^2}\right) \mathbf{E_S} = \frac{4\pi e}{c^2} \frac{\partial}{\partial t} (\mathbf{v_{Ie}}\delta n_e - \mathbf{v_{Ii}}\delta n_i),
\label{eq_waveeq}
\end{equation}
where $\mathbf{E_S}$ is the scattered electric field, $\mathbf{v_{Ie,i}}$ are the oscillation velocities in the incident electric field, and $\delta n_{e,i}$ are the density fluctuations of electrons and ions. The density fluctuations along $\mathbf{k}$ ($x$ direction) in Eq. \eqref{eq_waveeq} are obtained from a particle-in-cell (PIC) simulation. We performed 1D PIC simulations with the Smilei open source code.\cite{derouillat18cpc} The consistency of the results was confirmed with two independent open-source codes, Smilei \cite{derouillat18cpc} and EPOCH. \cite{arber15ppcf} The simulation domain was initially filled with the two components of the stationary and moving plasmas. The initial plasma conditions were the same as in Sec. \ref{sec_theory}. The following numerical parameters were used. The grid size was $\Delta x = \lambda_{d}$, where $\lambda_D$ is the Debye length. The time step was $\Delta t=0.9\Delta x/c$. The number of grid cells and time steps were $N_x = 32768$ and $N_t = 65536$, which correspond to a system size of $L_{x}\omega_{pe}/c\sim 1465$ and a computational time of $T\omega_{pe}\sim 2638$, respectively. The number of particles per cell was 4000. Periodic boundaries were applied at the ends of the simulation domain. The propagation angle of the incident electromagnetic wave was \SI{45}{\degree} relative to the $x$ axis. The perpendicular (to the $x$ axis) component of the wavenumber was assumed to be constant $\nabla_{\perp}^{2}=-k_{I\perp}^{2}=-k_{S\perp}^{2}$ in the scattering process. The subscript $\perp$ represents components perpendicular to the $\mathbf{k}$ ($x$) direction. With the above assumptions, Eq. \eqref{eq_waveeq} becomes $(-c^2\partial^2/\partial x^2 + \partial^2/\partial t^2+\omega_{p}^2 + c^2 k_{\perp}^2)\mathbf{E_S}=4\pi e [\partial(\mathbf{v_{Ie}}\delta n_e - \mathbf{v_{Ii}}\delta n_i)/\partial t]$. Since the intensity of the incident electromagnetic wave is weak enough to neglect the Lorentz force from the magnetic field, the oscillation velocities of the electrons and ions are given by $\mathbf{v_{Ie,i}}=(q_{e,i}/m_{e,i}) \mathbf{E_{I}}$, where $\mathbf{E_I}$ is the incident electric field. 

\begin{figure}
    \centering
    \includegraphics[clip,width=\hsize]{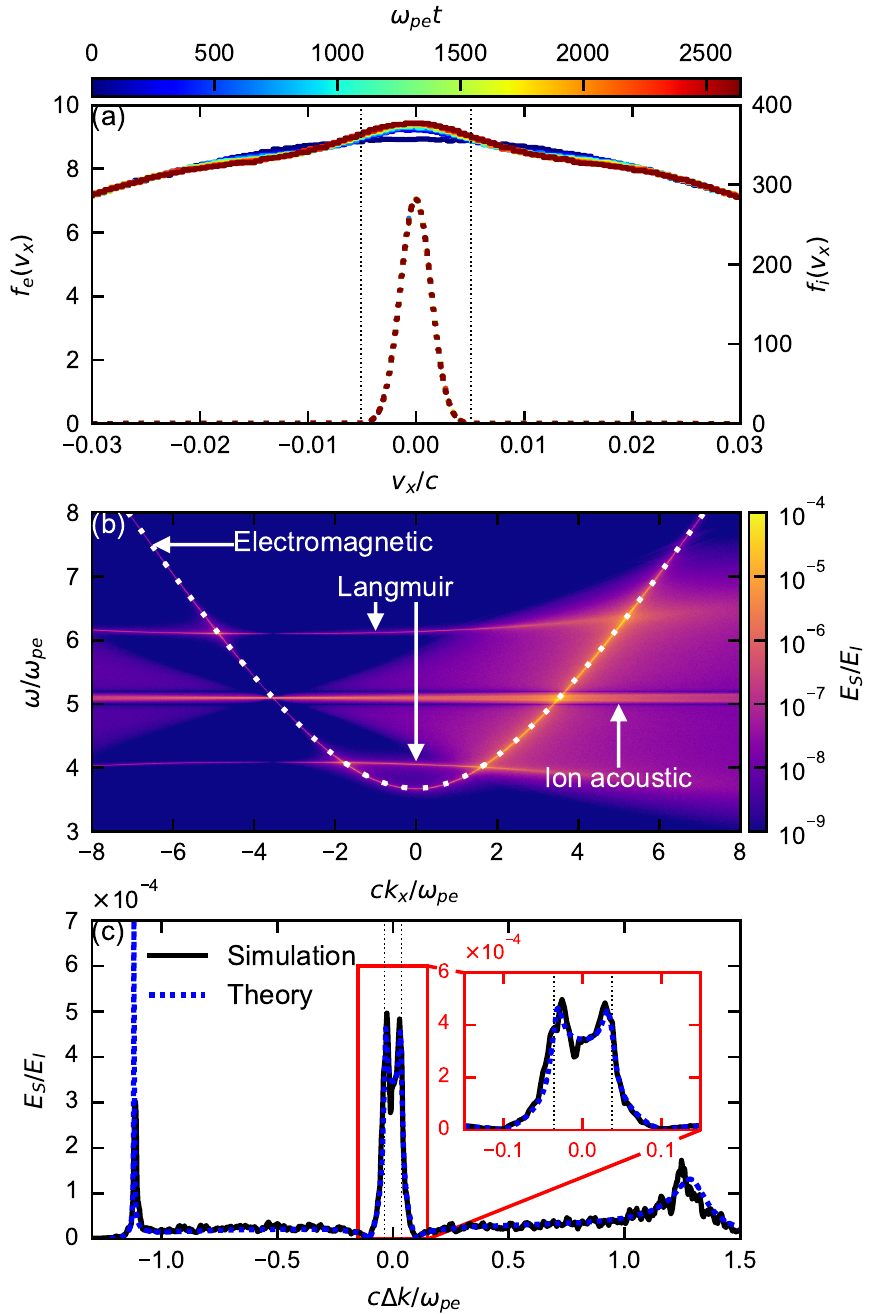}
    \caption{A simulated spectrum when $v_{d2}/v_{te} = 0$. (a) The temporal evolution of the distribution functions. (b) The dispersion relation of the scattered waves. (c) The CTS spectra.}
    \label{fig_3}
\end{figure}

Figure \ref{fig_3} shows the numerical results when $v_{d2}=0$. Figure \ref{fig_3}(a) shows the time evolution of the electron (solid) and ion (dotted) distribution functions. The color corresponds to the time. Since the electron temperature is higher than the ion temperature in the initial conditions, the electron and ion distribution functions slightly change over time to balance the ion and electron energies. The Fourier-transformed scattered electric field is shown in Fig. \ref{fig_3}(b). In the Fourier transformation, the Hann window is used along the temporal axis, $0.5-0.5\cos(2\pi t/T)$. Therefore, the dispersion relation in Fig. \ref{fig_3}(b) mainly reflects the time $t\sim T/2=1319\omega_{pe}^{-1}$. The incident wave corresponds to $(ck/\omega_{pe},\omega/\omega_{pe}) \sim (-3.5,5.1)$. There are three modes in the electric field: Langmuir waves at $\omega/\omega_{pe}\sim 4.1,6.1$, ion-acoustic waves at $\omega/\omega_{pe}\sim 5.1$, and electromagnetic waves indicated by the dotted white curve. Picking up the electric field on the dispersion relation of electromagnetic waves, we obtain the simulated CTS spectrum shown in Fig. \ref{fig_3}(c). The double peaks at $c\Delta k/\omega_{pe}\sim 0$ in the inset of Fig. \ref{fig_3}(c) correspond to the ion-acoustic feature, and the peaks at $c\Delta k/\omega_{pe}\sim -1.1$ and 1.3 to the electron feature. The simulated ion-acoustic spectrum shows symmetric double peaks in Fig. \ref{fig_1}(a), because the electron distribution function remains symmetric about $v=0$ and the Landau damping is nearly equal for the two peaks. The theoretical spectrum is shown by the blue dashed curve, using the distribution functions in Fig. \ref{fig_3}(a). The simulated spectrum of the electron and ion features agrees well with the theoretical ones. Note that the peak of the electron feature at $c\Delta k/\omega_{pe}\sim -1.1$ satisfies the resonant condition well and behaves like a delta function. Because the wavenumber resolution of the simulation is insufficient to resolve the peak, the peak intensities of the simulated and theoretical spectra seem different. 

\begin{figure}
    \centering
    \includegraphics[clip,width=\hsize]{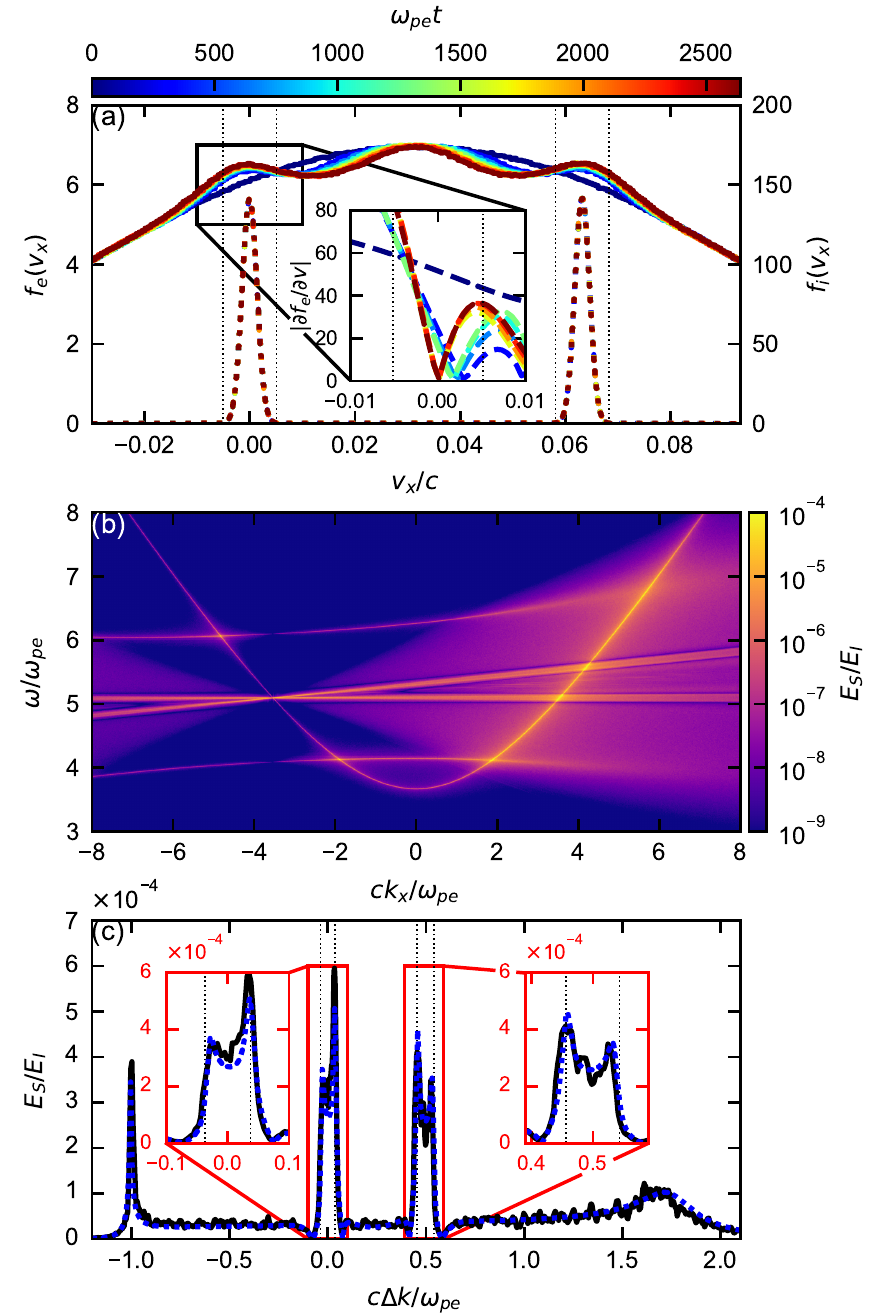}
    \caption{The simulated spectrum when $v_{d2}/v_{te} = 1$. (a) The temporal evolution of the distribution functions. (b) The dispersion relation of the scattered waves. (c) The CTS spectra.}
    \label{fig_4}
\end{figure}

Figure \ref{fig_4} shows the same plot as in Fig. \ref{fig_3} but for $v_{d2}/v_{te}=1$. The inset in Fig. \ref{fig_4}(a) shows the derivative of the electron distribution function, i.e., the relative rate of electron damping around the resonant phase velocities of the stationary plasma. The derivative shows a similar shape in both the stationary and moving plasmas. Since the simulated distribution function in Fig. \ref{fig_4}(a) contains noise, we fit the distribution function with two Maxwell distributions to reduce the noise and calculate the derivative in the inset. The ion distribution looks almost unchanged while the electron distribution varies with time. The derivative of the electron distribution function is smaller at the right peak in the stationary plasma and at the left peak in the moving plasma. Figures \ref{fig_4}(b) and \ref{fig_4}(c) show two ion-acoustic features relevant to the counterstreaming plasmas. The simulated spectrum is consistent with the theoretical one. As discussed in Sec. \ref{sec_theory}, the peaks with smaller derivatives of the electron distribution function are stronger than the others. 

\begin{figure}
    \centering
    \includegraphics[clip,width=\hsize]{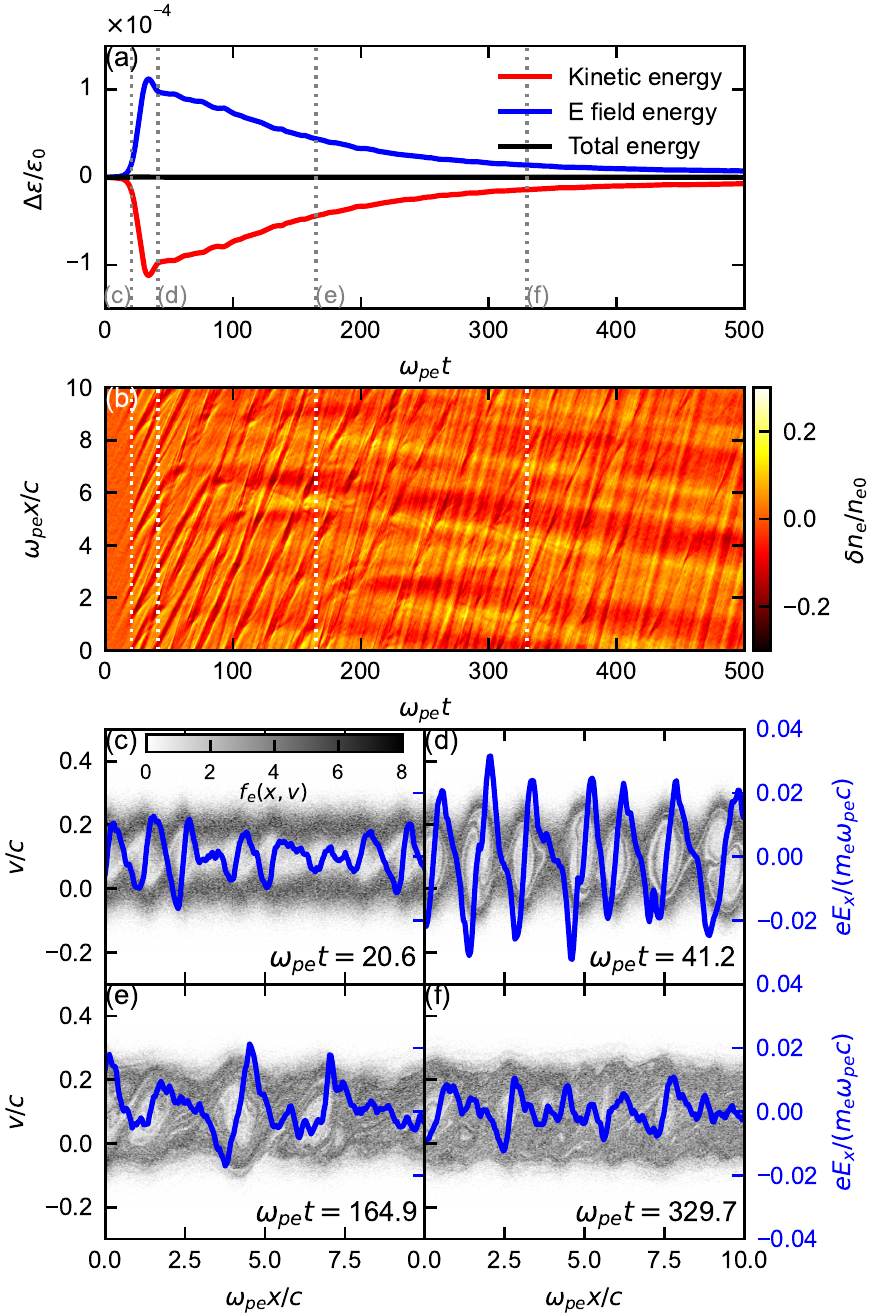}
    \caption{The growth of a two-stream-type instability when $v_{d2}/v_{te} = 3$. (a) The time evolution of the energy. (b) The electron density fluctuation in $x-t$ space. (c)--(f) The electron phase-space plot and electric field profile at $\omega_{pe}t=20.6$, 41.2, 164.9, and 329.7 [dotted lines in (a) and (b)].}
    \label{fig_5}
\end{figure}

In Fig. \ref{fig_5}(a), we plot the changes in the kinetic, electric, and total energies in the simulation domain as a function of time when $v_{d2}/v_{te}=3$. Note that the temporal axis is $\sim 5$ times shorter than the total computation time. The electric field energy increases at $\omega_{pe}t\lesssim 30$. The spatiotemporal evolution of the electron density fluctuation in Fig. \ref{fig_5}(b) shows the electrostatic wave propagation. The density fluctuation is normalized to the averaged density of $n_{e0}$. The wave amplitude grows to $\omega_{pe}t\lesssim 30$ with a phase velocity of $\sim v_{d2}/2$. This is consistent with the linear analysis of the dispersion relation in Fig. \ref{fig_2} and therefore, the wave growth is considered to be the result of a two-stream-type instability. The electron phase spaces at $\omega_{pe}t=20.6$, 41.2, 164.9, and 329.7 are shown in Figs. \ref{fig_5}(c)--(f), respectively. The electron distribution function locally changes from the initial conditions. The two-stream-type instability is seen to grow linearly in Fig. \ref{fig_5}(c) and to saturate in Fig. \ref{fig_5}(d). The wavelength of the excited wave is comparable to the mode with the maximum growth rate in Fig. \ref{fig_2}. After saturation, the nonlinear phase of the two-stream instability takes place. The phase-space holes attract each other and combine as shown in Figs. \ref{fig_5}(e) and \ref{fig_5}(f), resulting in electron thermalization. This timescale is shorter than the total computation time and a quasi-equilibrium plasma forms afterward. 
\begin{figure}
    \centering
    \includegraphics[clip,width=\hsize]{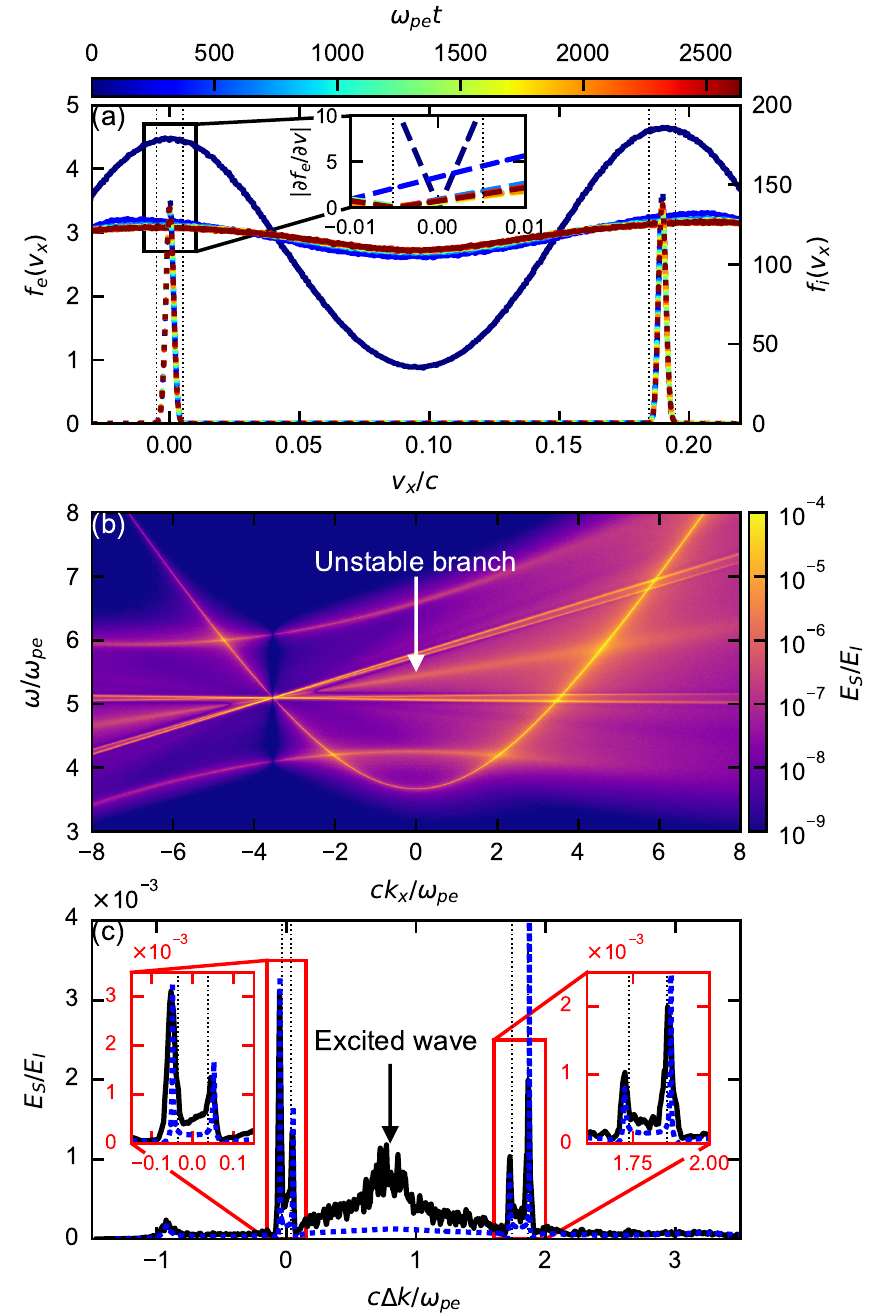}
    \caption{The simulated spectrum when $v_{d2}/v_{te} = 3$. (a) The temporal evolution of the distribution functions. (b) The dispersion relation of the scattered waves. (c) The CTS spectra.}
    \label{fig_6}
\end{figure}
Figure \ref{fig_6} shows the same plot as in Fig. \ref{fig_4} but with $v_{d2}/v_{te}=3$. As discussed above, the distribution function in Fig. \ref{fig_6}(a) changes over a short time and then stabilizes. In Fig. \ref{fig_6}(b), one can see a branch not found in Figs. \ref{fig_3}(b) and \ref{fig_4}(b), which is the unstable branch in Fig. \ref{fig_2}. In Fig. \ref{fig_6}(c), the mode of the excited wave is observed at $c\Delta k/\omega_{pe}\sim 0.8$. \cite{sakai20pop} Although the theoretical spectrum of the ion-acoustic feature at the initial timestep in Fig. \ref{fig_1}(c) shows symmetry, the simulated spectrum is asymmetric. The left peak in the left plasma and the right peak in the right plasma are larger than the other peaks, showing the opposite trend to the case when $v_{d2}/v_{te}=1$. The theoretical spectrum, including the time evolution of the distribution function as the blue curve, is consistent with the simulated ion-acoustic feature. As shown in the inset of Fig. \ref{fig_6}(a), while the derivative of the electron distribution function is equal for two peaks at the beginning of the simulation, the derivative at the right peak is larger than that at the left peak after growing the two-stream-type instability. Since the derivative of the electron distribution function is proportional to the rate of the electron Landau damping, the asymmetry in the ion-acoustic feature is considered to originate from the electron Landau damping. 

\section{Discussion and Summary}
\label{sec_discussion}

Comparing $S_{1+2}$ with $S_{1} + S_{2}$ in Fig. \ref{fig_1}, $S_{1+2}$ is different from $S_{1} + S_{2}$ when the distribution functions of the two plasmas overlap with each other near the phase velocities of the resonant ion-acoustic waves. When $v_{d2}\lesssim v_{ti}$, the electron and ion distribution functions overlap and the behavior of the dynamic structure factor can be complicated. When $v_{ti}\ll v_{d2}\lesssim v_{te}$, as shown in Fig. \ref{fig_1}(b), the electron distribution functions overlap but the ion distribution functions do not. In the second term on the right-hand side of Eq. \eqref{eq_form}, which describes the ion-acoustic feature, the difference between $S_{1+2}$ and $S_{1} + S_{2}$ is only the electron susceptibility. This relates to the derivative of the electron distribution function corresponding to the rate of electron Landau damping. The electron susceptibility has a minor effect on the resonant wavenumber and frequency because the ion susceptibility is much larger than the electron susceptibility and the ion susceptibility primarily determines the resonant condition. Therefore, the peak wavenumber of the CTS spectra is almost the same as in the single-plasma case and the peak intensities are determined by the derivative of the electron distribution function at the resonant velocity. This is consistent with the numerical result in Fig. \ref{fig_4}(c). In several Thomson scattering measurements in counterstreaming plasmas, e.g., Refs. \onlinecite{swadling20prl,ross12pop}, the CTS spectra showed asymmetric features. This is, therefore, a common feature of CTS spectra in counterstreaming plasmas. When $v_{d2}\gtrsim v_{te}$, as shown in Fig. \ref{fig_1}(c), there are no overlaps in the distribution functions and $S_{1+2}$ is the same as $S_{1}+S_{2}$. However, the electron distribution function is unstable and changes over a short time, as shown in Fig. \ref{fig_5}. The simulated spectrum in Fig. \ref{fig_6}(c) can be explained by the electron distribution function after the saturation of the two-stream-type instability. 

The typical plasma parameter in a laser-driven collisionless shock experiment, which is proportional to the ratio of the plasma frequency to the collision frequency, is $\Lambda=4\pi n_e\lambda_D \sim 10^{5-6}\gg 1$, \cite{kuramitsu16apj} i.e., the collision rarely perturbs the collective plasma oscillation. We have to be careful with low-frequency phenomena, however, as electrons are completely collisionless in electron-scale dynamics. If electrons were very collisional, CTS would not work. In many laser experiments,  CTS electron and ion features have shown excellent data so far, and both of these effects come from collisionless electron motion. 

The spectral asymmetry in the ion-acoustic feature of CTS is normally explained by the relative drift between electrons and ions or currents. \cite{froula11,sakai22srep} However, the spectral asymmetry arises in counterstreaming plasmas, as shown in Figs. \ref{fig_1}(b) and \ref{fig_4}(c). Even if the two components of the ion-acoustic feature are not observed experimentally due to the finite dynamic range of detectors and the low density compared to the main component, the overlap in the electron distribution functions can cause spectral asymmetry. Since the ion-acoustic feature itself measures the derivative of the electron distribution function at the resonant velocity, it is difficult to estimate the entire shape of the electron distribution function. In order to determine whether the spectral asymmetry in the ion-acoustic feature results from the net current or a non-Maxwellian distribution function, it is better to observe both the electron and ion features of CTS simultaneously. In the case of only an ion-acoustic feature, it is better to analyze CTS spectra using multiple-component plasmas \cite{sakai20pop} or to obtain angularly-resolved spectra to vary the resonant velocity. \cite{milder21prl,milder21pop}

When the velocity difference between the two components in counterstreaming plasmas exceeds the electron thermal velocity, the waves are excited because of the two-stream-type instability. When we observe CTS electron features, the spectrum shows a strong peak corresponding to the unstable mode. \cite{sakai20pop} In addition to this, the spectrum of the ion-acoustic feature is also modified due to the change in the derivative of the electron distribution function and shows a characteristic shape, as shown in Fig. \ref{fig_6}(c). By observing the electron feature showing the enhanced unstable mode and the ion-acoustic feature with the spectral asymmetry at the same time, one can conclude from the CTS data that a two-stream-type instability is taking place. This could be a new diagnostic with which to identify the instability in laboratory experiments. This measurement does not require very high spatial and/or temporal resolution because CTS measures the resonant wave amplitude. The timescale of the excited wave is much longer than that of the instability. Based on the simulation conducted here, the observable timescale is at least $T=2628 \omega_{pe}^{-1}\sim \SI{15}{ps}$ with an electron density of $10^{19}~\mathrm{cm^{-3}}$. Since the plasma state at the end of the simulation is almost steady, the timescale can be much longer. As the observable wavenumber and frequency with CTS are almost fixed by the optics (the wavelength of the incident electromagnetic field and the scattering angle), it is necessary to know which mode is excited and design an appropriate spectrometer. In Fig. \ref{fig_2}, the wavenumber with the maximum growth rate is $ck/\omega_{pe}\sim 5.6$, so we need to select the $\mathbf{k}$-vector to be around this value. When the electron density is ${10^{19}}~\mathrm{cm^{-3}}$ and the wavelength of the probe beam is \SI{1064} {nm}, a scattering angle of $\sim \SI{15}{\degree}$ is appropriate to observe the maximum growth mode. 

In this paper, we have considered the electron two-stream instability that is the fastest-growing mode in two-stream plasmas. After the saturation of the electron two-stream instability, the ion two-stream instability grows. \cite{kato10pop,ohira08apj} The instability generates an oblique mode propagating nearly perpendicular to the flow direction and the timescale of this instability is $\sim 100$ times larger than the electron two-stream instability. Our simulation employs 1D geometry and cannot treat the oblique mode. The effect of the ion two-stream instability in CTS will be reported in a future paper.

In summary, we have theoretically and numerically investigated the ion-acoustic features of collective Thomson scattering in two-stream plasmas. The theoretical spectrum from the two-plasma state is different from the sum of the individual spectra. When the velocity difference between the two components is approximately the electron thermal velocity, the rate of electron Landau damping is different for the two peaks due to the overlapped electron distribution functions and the CTS spectrum shows asymmetry. This is consistent with the numerical simulation. When the velocity difference exceeds the electron thermal velocity, a two-stream-type instability develops and the electron distribution function varies over time. The derivative of the distribution function becomes different for the two peaks and a spectral asymmetry arises that shows an opposite trend to the $v_{d2}/v_{te}=1$ case. Our results show that the non-Maxwellian distribution functions significantly affect the CTS spectra. Further investigations on CTS will enable us to measure the entire shape of the electron and ion distribution functions and their instabilities. 

\begin{acknowledgments}
This work was supported by JSPS KAKENHI grant numbers JP22H01195, JP22K14020, JP21J20499, JP20KK0064, JP19K21865, and the Sumitomo Foundation for environmental research projects (203099). 
Parts of this work were performed on the ``Plasma Simulator'' of the National Institute for Fusion Science with the support and under the auspices of the NIFS Collaboration Research program (NIFS22KISS028). 
\end{acknowledgments}

\appendix
\section{Dynamic structure factor with a realistic mass ratio}
\label{sec_massratio}

\begin{figure*}
    \centering
    \includegraphics[clip,width=\hsize]{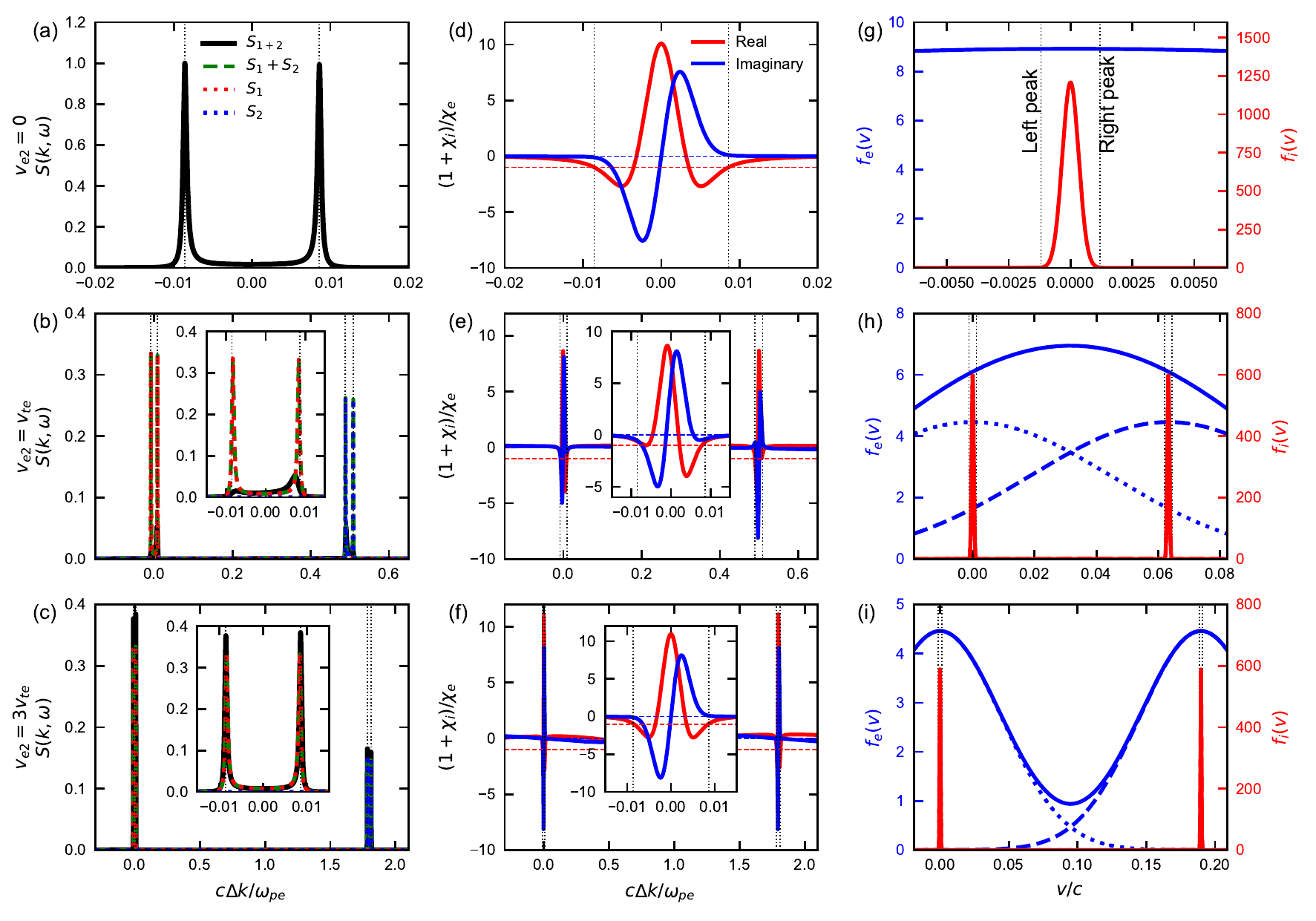}
    \caption{The dynamic structure factor with a realistic mass ratio ($m_i/m_e=1836$). (a)--(c) CTS spectra, (d)--(f) $(1+\chi_i)/\chi_e$, (g)--(i) Distribution functions.}
    \label{fig_7}
\end{figure*}

We set the mass ratio to 100 to reduce the computational resources required for the simulation in Sec. \ref{sec_sim}. While a numerical simulation with a realistic mass ratio requires a huge computation time and system size, the dynamic structure factor is easily calculated. We plot this in Fig. \ref{fig_7}, which is the same as Fig. \ref{fig_1} but with a realistic proton mass ($m_i/m_e=1836$). The spectral range is narrower than that with the reduced mass ratio because the thermal velocity of ions and the sound velocity are small. The shapes of $S_{1+2}$ are similar to those in Fig. \ref{fig_1}. This can be explained by the same strategy as in Sec. \ref{sec_theory}: the derivative of the electron distribution function at the resonant phase velocity.  

\bibliography{netsi}

\begin{thebibliography}{26}%
\makeatletter
\providecommand \@ifxundefined [1]{%
 \@ifx{#1\undefined}
}%
\providecommand \@ifnum [1]{%
 \ifnum #1\expandafter \@firstoftwo
 \else \expandafter \@secondoftwo
 \fi
}%
\providecommand \@ifx [1]{%
 \ifx #1\expandafter \@firstoftwo
 \else \expandafter \@secondoftwo
 \fi
}%
\providecommand \natexlab [1]{#1}%
\providecommand \enquote  [1]{``#1''}%
\providecommand \bibnamefont  [1]{#1}%
\providecommand \bibfnamefont [1]{#1}%
\providecommand \citenamefont [1]{#1}%
\providecommand \href@noop [0]{\@secondoftwo}%
\providecommand \href [0]{\begingroup \@sanitize@url \@href}%
\providecommand \@href[1]{\@@startlink{#1}\@@href}%
\providecommand \@@href[1]{\endgroup#1\@@endlink}%
\providecommand \@sanitize@url [0]{\catcode `\\12\catcode `\$12\catcode
  `\&12\catcode `\#12\catcode `\^12\catcode `\_12\catcode `\%12\relax}%
\providecommand \@@startlink[1]{}%
\providecommand \@@endlink[0]{}%
\providecommand \url  [0]{\begingroup\@sanitize@url \@url }%
\providecommand \@url [1]{\endgroup\@href {#1}{\urlprefix }}%
\providecommand \urlprefix  [0]{URL }%
\providecommand \Eprint [0]{\href }%
\providecommand \doibase [0]{https://doi.org/}%
\providecommand \selectlanguage [0]{\@gobble}%
\providecommand \bibinfo  [0]{\@secondoftwo}%
\providecommand \bibfield  [0]{\@secondoftwo}%
\providecommand \translation [1]{[#1]}%
\providecommand \BibitemOpen [0]{}%
\providecommand \bibitemStop [0]{}%
\providecommand \bibitemNoStop [0]{.\EOS\space}%
\providecommand \EOS [0]{\spacefactor3000\relax}%
\providecommand \BibitemShut  [1]{\csname bibitem#1\endcsname}%
\let\auto@bib@innerbib\@empty
\bibitem [{\citenamefont {Burgess}, \citenamefont {M{\"o}bius},\ and\
  \citenamefont {Scholer}(2012)}]{burgess12ssr}%
  \BibitemOpen
  \bibfield  {author} {\bibinfo {author} {\bibfnamefont {D.}~\bibnamefont
  {Burgess}}, \bibinfo {author} {\bibfnamefont {E.}~\bibnamefont
  {M{\"o}bius}},\ and\ \bibinfo {author} {\bibfnamefont {M.}~\bibnamefont
  {Scholer}},\ }\bibfield  {title} {\enquote {\bibinfo {title} {Ion
  acceleration at the earth’s bow shock},}\ }\href
  {https://doi.org/10.1007/s11214-012-9901-5} {\bibfield  {journal} {\bibinfo
  {journal} {Space Science Reviews}\ }\textbf {\bibinfo {volume} {173}},\
  \bibinfo {pages} {5--47} (\bibinfo {year} {2012})}\BibitemShut {NoStop}%
\bibitem [{\citenamefont {Matsukiyo}\ and\ \citenamefont
  {Scholer}(2006)}]{matsukiyo06jgr}%
  \BibitemOpen
  \bibfield  {author} {\bibinfo {author} {\bibfnamefont {S.}~\bibnamefont
  {Matsukiyo}}\ and\ \bibinfo {author} {\bibfnamefont {M.}~\bibnamefont
  {Scholer}},\ }\bibfield  {title} {\enquote {\bibinfo {title} {On
  microinstabilities in the foot of high mach number perpendicular shocks},}\
  }\href {https://doi.org/10.1029/2005JA011409} {\bibfield  {journal} {\bibinfo
   {journal} {Journal of Geophysical Research: Space Physics}\ }\textbf
  {\bibinfo {volume} {111}} (\bibinfo {year} {2006}),\
  10.1029/2005JA011409}\BibitemShut {NoStop}%
\bibitem [{\citenamefont {Umeda}\ \emph {et~al.}(2014)\citenamefont {Umeda},
  \citenamefont {Kidani}, \citenamefont {Matsukiyo},\ and\ \citenamefont
  {Yamazaki}}]{umeda14pop}%
  \BibitemOpen
  \bibfield  {author} {\bibinfo {author} {\bibfnamefont {T.}~\bibnamefont
  {Umeda}}, \bibinfo {author} {\bibfnamefont {Y.}~\bibnamefont {Kidani}},
  \bibinfo {author} {\bibfnamefont {S.}~\bibnamefont {Matsukiyo}},\ and\
  \bibinfo {author} {\bibfnamefont {R.}~\bibnamefont {Yamazaki}},\ }\bibfield
  {title} {\enquote {\bibinfo {title} {Dynamics and microinstabilities at
  perpendicular collisionless shock: A comparison of large-scale
  two-dimensional full particle simulations with different ion to electron mass
  ratio},}\ }\href {https://doi.org/10.1063/1.4863836} {\bibfield  {journal}
  {\bibinfo  {journal} {Physics of Plasmas}\ }\textbf {\bibinfo {volume}
  {21}},\ \bibinfo {pages} {022102} (\bibinfo {year} {2014})}\BibitemShut
  {NoStop}%
\bibitem [{\citenamefont {Lowe}\ and\ \citenamefont
  {Burgess}(2003)}]{lowe03angeo}%
  \BibitemOpen
  \bibfield  {author} {\bibinfo {author} {\bibfnamefont {R.~E.}\ \bibnamefont
  {Lowe}}\ and\ \bibinfo {author} {\bibfnamefont {D.}~\bibnamefont {Burgess}},\
  }\bibfield  {title} {\enquote {\bibinfo {title} {The properties and causes of
  rippling in quasi-perpendicular collisionless shock fronts},}\ }\href
  {https://doi.org/10.5194/angeo-21-671-2003} {\bibfield  {journal} {\bibinfo
  {journal} {Annales Geophysicae}\ }\textbf {\bibinfo {volume} {21}},\ \bibinfo
  {pages} {671--679} (\bibinfo {year} {2003})}\BibitemShut {NoStop}%
\bibitem [{\citenamefont {Takabe}\ and\ \citenamefont
  {Kuramitsu}(2021)}]{takabe21hpl}%
  \BibitemOpen
  \bibfield  {author} {\bibinfo {author} {\bibfnamefont {H.}~\bibnamefont
  {Takabe}}\ and\ \bibinfo {author} {\bibfnamefont {Y.}~\bibnamefont
  {Kuramitsu}},\ }\bibfield  {title} {\enquote {\bibinfo {title} {Recent
  progress of laboratory astrophysics with intense lasers},}\ }\href
  {https://doi.org/10.1017/hpl.2021.35} {\bibfield  {journal} {\bibinfo
  {journal} {High Power Laser Science and Engineering}\ }\textbf {\bibinfo
  {volume} {9}},\ \bibinfo {pages} {e49} (\bibinfo {year} {2021})}\BibitemShut
  {NoStop}%
\bibitem [{\citenamefont {Kuramitsu}\ \emph {et~al.}(2012)\citenamefont
  {Kuramitsu}, \citenamefont {Sakawa}, \citenamefont {Morita}, \citenamefont
  {Ide}, \citenamefont {Nishio}, \citenamefont {Tanji}, \citenamefont {Aoki},
  \citenamefont {Dono}, \citenamefont {Gregory}, \citenamefont {Waugh},
  \citenamefont {Woolsey}, \citenamefont {Dizi{\`{e}}re}, \citenamefont
  {Pelka}, \citenamefont {Ravasio}, \citenamefont {Loupias}, \citenamefont
  {Koenig}, \citenamefont {Pikuz}, \citenamefont {Li}, \citenamefont {Zhang},
  \citenamefont {Liu}, \citenamefont {Zhong}, \citenamefont {Zhang},
  \citenamefont {Gregori}, \citenamefont {Nakanii}, \citenamefont {Kondo},
  \citenamefont {Mori}, \citenamefont {Miura}, \citenamefont {Kodama},
  \citenamefont {Kitagawa}, \citenamefont {Mima}, \citenamefont {Tanaka},
  \citenamefont {Azechi}, \citenamefont {Moritaka}, \citenamefont {Matsumoto},
  \citenamefont {Sano}, \citenamefont {Mizuta}, \citenamefont {Ohnishi},
  \citenamefont {Hoshino},\ and\ \citenamefont {Takabe}}]{kuramitsu12ppcf}%
  \BibitemOpen
  \bibfield  {author} {\bibinfo {author} {\bibfnamefont {Y.}~\bibnamefont
  {Kuramitsu}}, \bibinfo {author} {\bibfnamefont {Y.}~\bibnamefont {Sakawa}},
  \bibinfo {author} {\bibfnamefont {T.}~\bibnamefont {Morita}}, \bibinfo
  {author} {\bibfnamefont {T.}~\bibnamefont {Ide}}, \bibinfo {author}
  {\bibfnamefont {K.}~\bibnamefont {Nishio}}, \bibinfo {author} {\bibfnamefont
  {H.}~\bibnamefont {Tanji}}, \bibinfo {author} {\bibfnamefont
  {H.}~\bibnamefont {Aoki}}, \bibinfo {author} {\bibfnamefont {S.}~\bibnamefont
  {Dono}}, \bibinfo {author} {\bibfnamefont {C.~D.}\ \bibnamefont {Gregory}},
  \bibinfo {author} {\bibfnamefont {J.~N.}\ \bibnamefont {Waugh}}, \bibinfo
  {author} {\bibfnamefont {N.}~\bibnamefont {Woolsey}}, \bibinfo {author}
  {\bibfnamefont {A.}~\bibnamefont {Dizi{\`{e}}re}}, \bibinfo {author}
  {\bibfnamefont {A.}~\bibnamefont {Pelka}}, \bibinfo {author} {\bibfnamefont
  {A.}~\bibnamefont {Ravasio}}, \bibinfo {author} {\bibfnamefont
  {B.}~\bibnamefont {Loupias}}, \bibinfo {author} {\bibfnamefont
  {M.}~\bibnamefont {Koenig}}, \bibinfo {author} {\bibfnamefont {S.~A.}\
  \bibnamefont {Pikuz}}, \bibinfo {author} {\bibfnamefont {Y.~T.}\ \bibnamefont
  {Li}}, \bibinfo {author} {\bibfnamefont {Y.}~\bibnamefont {Zhang}}, \bibinfo
  {author} {\bibfnamefont {X.}~\bibnamefont {Liu}}, \bibinfo {author}
  {\bibfnamefont {J.~Y.}\ \bibnamefont {Zhong}}, \bibinfo {author}
  {\bibfnamefont {J.}~\bibnamefont {Zhang}}, \bibinfo {author} {\bibfnamefont
  {G.}~\bibnamefont {Gregori}}, \bibinfo {author} {\bibfnamefont
  {N.}~\bibnamefont {Nakanii}}, \bibinfo {author} {\bibfnamefont
  {K.}~\bibnamefont {Kondo}}, \bibinfo {author} {\bibfnamefont
  {Y.}~\bibnamefont {Mori}}, \bibinfo {author} {\bibfnamefont {E.}~\bibnamefont
  {Miura}}, \bibinfo {author} {\bibfnamefont {R.}~\bibnamefont {Kodama}},
  \bibinfo {author} {\bibfnamefont {Y.}~\bibnamefont {Kitagawa}}, \bibinfo
  {author} {\bibfnamefont {K.}~\bibnamefont {Mima}}, \bibinfo {author}
  {\bibfnamefont {K.~A.}\ \bibnamefont {Tanaka}}, \bibinfo {author}
  {\bibfnamefont {H.}~\bibnamefont {Azechi}}, \bibinfo {author} {\bibfnamefont
  {T.}~\bibnamefont {Moritaka}}, \bibinfo {author} {\bibfnamefont
  {Y.}~\bibnamefont {Matsumoto}}, \bibinfo {author} {\bibfnamefont
  {T.}~\bibnamefont {Sano}}, \bibinfo {author} {\bibfnamefont {A.}~\bibnamefont
  {Mizuta}}, \bibinfo {author} {\bibfnamefont {N.}~\bibnamefont {Ohnishi}},
  \bibinfo {author} {\bibfnamefont {M.}~\bibnamefont {Hoshino}},\ and\ \bibinfo
  {author} {\bibfnamefont {H.}~\bibnamefont {Takabe}},\ }\bibfield  {title}
  {\enquote {\bibinfo {title} {Laboratory investigations on the origins of
  cosmic rays},}\ }\href {https://doi.org/10.1088/0741-3335/54/12/124049}
  {\bibfield  {journal} {\bibinfo  {journal} {Plasma Physics and Controlled
  Fusion}\ }\textbf {\bibinfo {volume} {54}},\ \bibinfo {pages} {124049}
  (\bibinfo {year} {2012})}\BibitemShut {NoStop}%
\bibitem [{\citenamefont {Froula}\ \emph {et~al.}(2011)\citenamefont {Froula},
  \citenamefont {Glenzer}, \citenamefont {Luhmann},\ and\ \citenamefont
  {Sheffield}}]{froula11}%
  \BibitemOpen
  \bibfield  {author} {\bibinfo {author} {\bibfnamefont {D.~H.}\ \bibnamefont
  {Froula}}, \bibinfo {author} {\bibfnamefont {S.~H.}\ \bibnamefont {Glenzer}},
  \bibinfo {author} {\bibfnamefont {N.~C.}\ \bibnamefont {Luhmann}},\ and\
  \bibinfo {author} {\bibfnamefont {J.}~\bibnamefont {Sheffield}},\ }\href
  {https://doi.org/10.1016/C2009-0-20048-1} {\emph {\bibinfo {title} {Plasma
  Scattering of Electromagnetic Radiation: Theory and Measurement
  Techniques}}},\ \bibinfo {edition} {2nd}\ ed.\ (\bibinfo  {publisher}
  {Academic Press},\ \bibinfo {address} {Amsterdam},\ \bibinfo {year}
  {2011})\BibitemShut {NoStop}%
\bibitem [{\citenamefont {Milder}\ \emph
  {et~al.}(2021{\natexlab{a}})\citenamefont {Milder}, \citenamefont {Katz},
  \citenamefont {Boni}, \citenamefont {Palastro}, \citenamefont {Sherlock},
  \citenamefont {Rozmus},\ and\ \citenamefont {Froula}}]{milder21prl}%
  \BibitemOpen
  \bibfield  {author} {\bibinfo {author} {\bibfnamefont {A.~L.}\ \bibnamefont
  {Milder}}, \bibinfo {author} {\bibfnamefont {J.}~\bibnamefont {Katz}},
  \bibinfo {author} {\bibfnamefont {R.}~\bibnamefont {Boni}}, \bibinfo {author}
  {\bibfnamefont {J.~P.}\ \bibnamefont {Palastro}}, \bibinfo {author}
  {\bibfnamefont {M.}~\bibnamefont {Sherlock}}, \bibinfo {author}
  {\bibfnamefont {W.}~\bibnamefont {Rozmus}},\ and\ \bibinfo {author}
  {\bibfnamefont {D.~H.}\ \bibnamefont {Froula}},\ }\bibfield  {title}
  {\enquote {\bibinfo {title} {Measurements of non-maxwellian electron
  distribution functions and their effect on laser heating},}\ }\href
  {https://doi.org/10.1103/PhysRevLett.127.015001} {\bibfield  {journal}
  {\bibinfo  {journal} {Phys. Rev. Lett.}\ }\textbf {\bibinfo {volume} {127}},\
  \bibinfo {pages} {015001} (\bibinfo {year} {2021}{\natexlab{a}})}\BibitemShut
  {NoStop}%
\bibitem [{\citenamefont {Milder}\ \emph
  {et~al.}(2021{\natexlab{b}})\citenamefont {Milder}, \citenamefont {Katz},
  \citenamefont {Boni}, \citenamefont {Palastro}, \citenamefont {Sherlock},
  \citenamefont {Rozmus},\ and\ \citenamefont {Froula}}]{milder21pop}%
  \BibitemOpen
  \bibfield  {author} {\bibinfo {author} {\bibfnamefont {A.~L.}\ \bibnamefont
  {Milder}}, \bibinfo {author} {\bibfnamefont {J.}~\bibnamefont {Katz}},
  \bibinfo {author} {\bibfnamefont {R.}~\bibnamefont {Boni}}, \bibinfo {author}
  {\bibfnamefont {J.~P.}\ \bibnamefont {Palastro}}, \bibinfo {author}
  {\bibfnamefont {M.}~\bibnamefont {Sherlock}}, \bibinfo {author}
  {\bibfnamefont {W.}~\bibnamefont {Rozmus}},\ and\ \bibinfo {author}
  {\bibfnamefont {D.~H.}\ \bibnamefont {Froula}},\ }\bibfield  {title}
  {\enquote {\bibinfo {title} {Statistical analysis of non-maxwellian electron
  distribution functions measured with angularly resolved thomson
  scattering},}\ }\href {https://doi.org/10.1063/5.0041504} {\bibfield
  {journal} {\bibinfo  {journal} {Physics of Plasmas}\ }\textbf {\bibinfo
  {volume} {28}},\ \bibinfo {pages} {082102} (\bibinfo {year}
  {2021}{\natexlab{b}})}\BibitemShut {NoStop}%
\bibitem [{\citenamefont {Turnbull}\ \emph {et~al.}(2020)\citenamefont
  {Turnbull}, \citenamefont {Colaïtis}, \citenamefont {Hansen}, \citenamefont
  {Milder}, \citenamefont {Palastro}, \citenamefont {Katz}, \citenamefont
  {Dorrer}, \citenamefont {Kruschwitz}, \citenamefont {Strozzi},\ and\
  \citenamefont {Froula}}]{turnbull20nphys}%
  \BibitemOpen
  \bibfield  {author} {\bibinfo {author} {\bibfnamefont {D.}~\bibnamefont
  {Turnbull}}, \bibinfo {author} {\bibfnamefont {A.}~\bibnamefont {Colaïtis}},
  \bibinfo {author} {\bibfnamefont {A.~M.}\ \bibnamefont {Hansen}}, \bibinfo
  {author} {\bibfnamefont {A.~L.}\ \bibnamefont {Milder}}, \bibinfo {author}
  {\bibfnamefont {J.~P.}\ \bibnamefont {Palastro}}, \bibinfo {author}
  {\bibfnamefont {J.}~\bibnamefont {Katz}}, \bibinfo {author} {\bibfnamefont
  {C.}~\bibnamefont {Dorrer}}, \bibinfo {author} {\bibfnamefont {B.~E.}\
  \bibnamefont {Kruschwitz}}, \bibinfo {author} {\bibfnamefont {D.~J.}\
  \bibnamefont {Strozzi}},\ and\ \bibinfo {author} {\bibfnamefont {D.~H.}\
  \bibnamefont {Froula}},\ }\bibfield  {title} {\enquote {\bibinfo {title}
  {Impact of the langdon effect on crossed-beam energy transfer},}\ }\href
  {https://doi.org/10.1038/s41567-019-0725-z} {\bibfield  {journal} {\bibinfo
  {journal} {Nature Physics}\ }\textbf {\bibinfo {volume} {16}} (\bibinfo
  {year} {2020}),\ 10.1038/s41567-019-0725-z}\BibitemShut {NoStop}%
\bibitem [{\citenamefont {Henchen}\ \emph {et~al.}(2018)\citenamefont
  {Henchen}, \citenamefont {Sherlock}, \citenamefont {Rozmus}, \citenamefont
  {Katz}, \citenamefont {Cao}, \citenamefont {Palastro},\ and\ \citenamefont
  {Froula}}]{henchen18prl}%
  \BibitemOpen
  \bibfield  {author} {\bibinfo {author} {\bibfnamefont {R.~J.}\ \bibnamefont
  {Henchen}}, \bibinfo {author} {\bibfnamefont {M.}~\bibnamefont {Sherlock}},
  \bibinfo {author} {\bibfnamefont {W.}~\bibnamefont {Rozmus}}, \bibinfo
  {author} {\bibfnamefont {J.}~\bibnamefont {Katz}}, \bibinfo {author}
  {\bibfnamefont {D.}~\bibnamefont {Cao}}, \bibinfo {author} {\bibfnamefont
  {J.~P.}\ \bibnamefont {Palastro}},\ and\ \bibinfo {author} {\bibfnamefont
  {D.~H.}\ \bibnamefont {Froula}},\ }\bibfield  {title} {\enquote {\bibinfo
  {title} {Observation of nonlocal heat flux using thomson scattering},}\
  }\href {https://doi.org/10.1103/PhysRevLett.121.125001} {\bibfield  {journal}
  {\bibinfo  {journal} {Phys. Rev. Lett.}\ }\textbf {\bibinfo {volume} {121}},\
  \bibinfo {pages} {125001} (\bibinfo {year} {2018})}\BibitemShut {NoStop}%
\bibitem [{\citenamefont {Sakai}\ \emph {et~al.}(2020)\citenamefont {Sakai},
  \citenamefont {Isayama}, \citenamefont {Bolouki}, \citenamefont {Habibi},
  \citenamefont {Liu}, \citenamefont {Hsieh}, \citenamefont {Chu},
  \citenamefont {Wang}, \citenamefont {Chen}, \citenamefont {Morita},
  \citenamefont {Tomita}, \citenamefont {Yamazaki}, \citenamefont {Sakawa},
  \citenamefont {Matsukiyo},\ and\ \citenamefont {Kuramitsu}}]{sakai20pop}%
  \BibitemOpen
  \bibfield  {author} {\bibinfo {author} {\bibfnamefont {K.}~\bibnamefont
  {Sakai}}, \bibinfo {author} {\bibfnamefont {S.}~\bibnamefont {Isayama}},
  \bibinfo {author} {\bibfnamefont {N.}~\bibnamefont {Bolouki}}, \bibinfo
  {author} {\bibfnamefont {M.~S.}\ \bibnamefont {Habibi}}, \bibinfo {author}
  {\bibfnamefont {Y.~L.}\ \bibnamefont {Liu}}, \bibinfo {author} {\bibfnamefont
  {Y.~H.}\ \bibnamefont {Hsieh}}, \bibinfo {author} {\bibfnamefont {H.~H.}\
  \bibnamefont {Chu}}, \bibinfo {author} {\bibfnamefont {J.}~\bibnamefont
  {Wang}}, \bibinfo {author} {\bibfnamefont {S.~H.}\ \bibnamefont {Chen}},
  \bibinfo {author} {\bibfnamefont {T.}~\bibnamefont {Morita}}, \bibinfo
  {author} {\bibfnamefont {K.}~\bibnamefont {Tomita}}, \bibinfo {author}
  {\bibfnamefont {R.}~\bibnamefont {Yamazaki}}, \bibinfo {author}
  {\bibfnamefont {Y.}~\bibnamefont {Sakawa}}, \bibinfo {author} {\bibfnamefont
  {S.}~\bibnamefont {Matsukiyo}},\ and\ \bibinfo {author} {\bibfnamefont
  {Y.}~\bibnamefont {Kuramitsu}},\ }\bibfield  {title} {\enquote {\bibinfo
  {title} {Collective thomson scattering in non-equilibrium laser produced
  two-stream plasmas},}\ }\href {https://doi.org/10.1063/5.0011935} {\bibfield
  {journal} {\bibinfo  {journal} {Physics of Plasmas}\ }\textbf {\bibinfo
  {volume} {27}},\ \bibinfo {pages} {103104} (\bibinfo {year}
  {2020})}\BibitemShut {NoStop}%
\bibitem [{\citenamefont {Matsukiyo}, \citenamefont {Kuramitsu},\ and\
  \citenamefont {Tomita}(2016)}]{matsukiyo16jpcs}%
  \BibitemOpen
  \bibfield  {author} {\bibinfo {author} {\bibfnamefont {S.}~\bibnamefont
  {Matsukiyo}}, \bibinfo {author} {\bibfnamefont {Y.}~\bibnamefont
  {Kuramitsu}},\ and\ \bibinfo {author} {\bibfnamefont {K.}~\bibnamefont
  {Tomita}},\ }\bibfield  {title} {\enquote {\bibinfo {title} {Collective
  scattering of an incident monochromatic circularly polarized wave in an
  unmagnetized non-equilibrium plasma},}\ }\href
  {https://doi.org/10.1088/1742-6596/688/1/012062} {\bibfield  {journal}
  {\bibinfo  {journal} {Journal of Physics: Conference Series}\ }\textbf
  {\bibinfo {volume} {688}},\ \bibinfo {pages} {012062} (\bibinfo {year}
  {2016})}\BibitemShut {NoStop}%
\bibitem [{\citenamefont {Saito}\ \emph {et~al.}(2000)\citenamefont {Saito},
  \citenamefont {Forme}, \citenamefont {Buchert}, \citenamefont {Nozawa},\ and\
  \citenamefont {Fujii}}]{saito00angeo}%
  \BibitemOpen
  \bibfield  {author} {\bibinfo {author} {\bibfnamefont {S.}~\bibnamefont
  {Saito}}, \bibinfo {author} {\bibfnamefont {F.~R.~E.}\ \bibnamefont {Forme}},
  \bibinfo {author} {\bibfnamefont {S.~C.}\ \bibnamefont {Buchert}}, \bibinfo
  {author} {\bibfnamefont {S.}~\bibnamefont {Nozawa}},\ and\ \bibinfo {author}
  {\bibfnamefont {R.}~\bibnamefont {Fujii}},\ }\bibfield  {title} {\enquote
  {\bibinfo {title} {Effects of a kappa distribution function of electrons on
  incoherent scatter spectra},}\ }\href
  {https://doi.org/10.1007/s00585-000-1216-2} {\bibfield  {journal} {\bibinfo
  {journal} {Annales Geophysicae}\ }\textbf {\bibinfo {volume} {18}},\ \bibinfo
  {pages} {1216--1223} (\bibinfo {year} {2000})}\BibitemShut {NoStop}%
\bibitem [{\citenamefont {Yamazaki}\ \emph {et~al.}(2022)\citenamefont
  {Yamazaki}, \citenamefont {Matsukiyo}, \citenamefont {Morita}, \citenamefont
  {Tanaka}, \citenamefont {Umeda}, \citenamefont {Aihara}, \citenamefont
  {Edamoto}, \citenamefont {Egashira}, \citenamefont {Hatsuyama}, \citenamefont
  {Higuchi}, \citenamefont {Hihara}, \citenamefont {Horie}, \citenamefont
  {Hoshino}, \citenamefont {Ishii}, \citenamefont {Ishizaka}, \citenamefont
  {Itadani}, \citenamefont {Izumi}, \citenamefont {Kambayashi}, \citenamefont
  {Kakuchi}, \citenamefont {Katsuki}, \citenamefont {Kawamura}, \citenamefont
  {Kawamura}, \citenamefont {Kisaka}, \citenamefont {Kojima}, \citenamefont
  {Konuma}, \citenamefont {Kumar}, \citenamefont {Minami}, \citenamefont
  {Miyata}, \citenamefont {Moritaka}, \citenamefont {Murakami}, \citenamefont
  {Nagashima}, \citenamefont {Nakagawa}, \citenamefont {Nishimoto},
  \citenamefont {Nishioka}, \citenamefont {Ohira}, \citenamefont {Ohnishi},
  \citenamefont {Ota}, \citenamefont {Ozaki}, \citenamefont {Sano},
  \citenamefont {Sakai}, \citenamefont {Sei}, \citenamefont {Shiota},
  \citenamefont {Shoji}, \citenamefont {Sugiyama}, \citenamefont {Suzuki},
  \citenamefont {Takagi}, \citenamefont {Toda}, \citenamefont {Tomita},
  \citenamefont {Tomiya}, \citenamefont {Yoneda}, \citenamefont {Takezaki},
  \citenamefont {Tomita}, \citenamefont {Kuramitsu},\ and\ \citenamefont
  {Sakawa}}]{yamazaki22pre}%
  \BibitemOpen
  \bibfield  {author} {\bibinfo {author} {\bibfnamefont {R.}~\bibnamefont
  {Yamazaki}}, \bibinfo {author} {\bibfnamefont {S.}~\bibnamefont {Matsukiyo}},
  \bibinfo {author} {\bibfnamefont {T.}~\bibnamefont {Morita}}, \bibinfo
  {author} {\bibfnamefont {S.~J.}\ \bibnamefont {Tanaka}}, \bibinfo {author}
  {\bibfnamefont {T.}~\bibnamefont {Umeda}}, \bibinfo {author} {\bibfnamefont
  {K.}~\bibnamefont {Aihara}}, \bibinfo {author} {\bibfnamefont
  {M.}~\bibnamefont {Edamoto}}, \bibinfo {author} {\bibfnamefont
  {S.}~\bibnamefont {Egashira}}, \bibinfo {author} {\bibfnamefont
  {R.}~\bibnamefont {Hatsuyama}}, \bibinfo {author} {\bibfnamefont
  {T.}~\bibnamefont {Higuchi}}, \bibinfo {author} {\bibfnamefont
  {T.}~\bibnamefont {Hihara}}, \bibinfo {author} {\bibfnamefont
  {Y.}~\bibnamefont {Horie}}, \bibinfo {author} {\bibfnamefont
  {M.}~\bibnamefont {Hoshino}}, \bibinfo {author} {\bibfnamefont
  {A.}~\bibnamefont {Ishii}}, \bibinfo {author} {\bibfnamefont
  {N.}~\bibnamefont {Ishizaka}}, \bibinfo {author} {\bibfnamefont
  {Y.}~\bibnamefont {Itadani}}, \bibinfo {author} {\bibfnamefont
  {T.}~\bibnamefont {Izumi}}, \bibinfo {author} {\bibfnamefont
  {S.}~\bibnamefont {Kambayashi}}, \bibinfo {author} {\bibfnamefont
  {S.}~\bibnamefont {Kakuchi}}, \bibinfo {author} {\bibfnamefont
  {N.}~\bibnamefont {Katsuki}}, \bibinfo {author} {\bibfnamefont
  {R.}~\bibnamefont {Kawamura}}, \bibinfo {author} {\bibfnamefont
  {Y.}~\bibnamefont {Kawamura}}, \bibinfo {author} {\bibfnamefont
  {S.}~\bibnamefont {Kisaka}}, \bibinfo {author} {\bibfnamefont
  {T.}~\bibnamefont {Kojima}}, \bibinfo {author} {\bibfnamefont
  {A.}~\bibnamefont {Konuma}}, \bibinfo {author} {\bibfnamefont
  {R.}~\bibnamefont {Kumar}}, \bibinfo {author} {\bibfnamefont
  {T.}~\bibnamefont {Minami}}, \bibinfo {author} {\bibfnamefont
  {I.}~\bibnamefont {Miyata}}, \bibinfo {author} {\bibfnamefont
  {T.}~\bibnamefont {Moritaka}}, \bibinfo {author} {\bibfnamefont
  {Y.}~\bibnamefont {Murakami}}, \bibinfo {author} {\bibfnamefont
  {K.}~\bibnamefont {Nagashima}}, \bibinfo {author} {\bibfnamefont
  {Y.}~\bibnamefont {Nakagawa}}, \bibinfo {author} {\bibfnamefont
  {T.}~\bibnamefont {Nishimoto}}, \bibinfo {author} {\bibfnamefont
  {Y.}~\bibnamefont {Nishioka}}, \bibinfo {author} {\bibfnamefont
  {Y.}~\bibnamefont {Ohira}}, \bibinfo {author} {\bibfnamefont
  {N.}~\bibnamefont {Ohnishi}}, \bibinfo {author} {\bibfnamefont
  {M.}~\bibnamefont {Ota}}, \bibinfo {author} {\bibfnamefont {N.}~\bibnamefont
  {Ozaki}}, \bibinfo {author} {\bibfnamefont {T.}~\bibnamefont {Sano}},
  \bibinfo {author} {\bibfnamefont {K.}~\bibnamefont {Sakai}}, \bibinfo
  {author} {\bibfnamefont {S.}~\bibnamefont {Sei}}, \bibinfo {author}
  {\bibfnamefont {J.}~\bibnamefont {Shiota}}, \bibinfo {author} {\bibfnamefont
  {Y.}~\bibnamefont {Shoji}}, \bibinfo {author} {\bibfnamefont
  {K.}~\bibnamefont {Sugiyama}}, \bibinfo {author} {\bibfnamefont
  {D.}~\bibnamefont {Suzuki}}, \bibinfo {author} {\bibfnamefont
  {M.}~\bibnamefont {Takagi}}, \bibinfo {author} {\bibfnamefont
  {H.}~\bibnamefont {Toda}}, \bibinfo {author} {\bibfnamefont {S.}~\bibnamefont
  {Tomita}}, \bibinfo {author} {\bibfnamefont {S.}~\bibnamefont {Tomiya}},
  \bibinfo {author} {\bibfnamefont {H.}~\bibnamefont {Yoneda}}, \bibinfo
  {author} {\bibfnamefont {T.}~\bibnamefont {Takezaki}}, \bibinfo {author}
  {\bibfnamefont {K.}~\bibnamefont {Tomita}}, \bibinfo {author} {\bibfnamefont
  {Y.}~\bibnamefont {Kuramitsu}},\ and\ \bibinfo {author} {\bibfnamefont
  {Y.}~\bibnamefont {Sakawa}},\ }\bibfield  {title} {\enquote {\bibinfo {title}
  {High-power laser experiment forming a supercritical collisionless shock in a
  magnetized uniform plasma at rest},}\ }\href
  {https://doi.org/10.1103/PhysRevE.105.025203} {\bibfield  {journal} {\bibinfo
   {journal} {Phys. Rev. E}\ }\textbf {\bibinfo {volume} {105}},\ \bibinfo
  {pages} {025203} (\bibinfo {year} {2022})}\BibitemShut {NoStop}%
\bibitem [{\citenamefont {Schaeffer}\ \emph {et~al.}(2019)\citenamefont
  {Schaeffer}, \citenamefont {Fox}, \citenamefont {Follett}, \citenamefont
  {Fiksel}, \citenamefont {Li}, \citenamefont {Matteucci}, \citenamefont
  {Bhattacharjee},\ and\ \citenamefont {Germaschewski}}]{schaeffer19prl}%
  \BibitemOpen
  \bibfield  {author} {\bibinfo {author} {\bibfnamefont {D.~B.}\ \bibnamefont
  {Schaeffer}}, \bibinfo {author} {\bibfnamefont {W.}~\bibnamefont {Fox}},
  \bibinfo {author} {\bibfnamefont {R.~K.}\ \bibnamefont {Follett}}, \bibinfo
  {author} {\bibfnamefont {G.}~\bibnamefont {Fiksel}}, \bibinfo {author}
  {\bibfnamefont {C.~K.}\ \bibnamefont {Li}}, \bibinfo {author} {\bibfnamefont
  {J.}~\bibnamefont {Matteucci}}, \bibinfo {author} {\bibfnamefont
  {A.}~\bibnamefont {Bhattacharjee}},\ and\ \bibinfo {author} {\bibfnamefont
  {K.}~\bibnamefont {Germaschewski}},\ }\bibfield  {title} {\enquote {\bibinfo
  {title} {Direct observations of particle dynamics in magnetized collisionless
  shock precursors in laser-produced plasmas},}\ }\href
  {https://doi.org/10.1103/PhysRevLett.122.245001} {\bibfield  {journal}
  {\bibinfo  {journal} {Phys. Rev. Lett.}\ }\textbf {\bibinfo {volume} {122}},\
  \bibinfo {pages} {245001} (\bibinfo {year} {2019})}\BibitemShut {NoStop}%
\bibitem [{\citenamefont {Swadling}\ \emph {et~al.}(2020)\citenamefont
  {Swadling}, \citenamefont {Bruulsema}, \citenamefont {Fiuza}, \citenamefont
  {Higginson}, \citenamefont {Huntington}, \citenamefont {Park}, \citenamefont
  {Pollock}, \citenamefont {Rozmus}, \citenamefont {Rinderknecht},
  \citenamefont {Katz}, \citenamefont {Birkel},\ and\ \citenamefont
  {Ross}}]{swadling20prl}%
  \BibitemOpen
  \bibfield  {author} {\bibinfo {author} {\bibfnamefont {G.~F.}\ \bibnamefont
  {Swadling}}, \bibinfo {author} {\bibfnamefont {C.}~\bibnamefont {Bruulsema}},
  \bibinfo {author} {\bibfnamefont {F.}~\bibnamefont {Fiuza}}, \bibinfo
  {author} {\bibfnamefont {D.~P.}\ \bibnamefont {Higginson}}, \bibinfo {author}
  {\bibfnamefont {C.~M.}\ \bibnamefont {Huntington}}, \bibinfo {author}
  {\bibfnamefont {H.-S.}\ \bibnamefont {Park}}, \bibinfo {author}
  {\bibfnamefont {B.~B.}\ \bibnamefont {Pollock}}, \bibinfo {author}
  {\bibfnamefont {W.}~\bibnamefont {Rozmus}}, \bibinfo {author} {\bibfnamefont
  {H.~G.}\ \bibnamefont {Rinderknecht}}, \bibinfo {author} {\bibfnamefont
  {J.}~\bibnamefont {Katz}}, \bibinfo {author} {\bibfnamefont {A.}~\bibnamefont
  {Birkel}},\ and\ \bibinfo {author} {\bibfnamefont {J.~S.}\ \bibnamefont
  {Ross}},\ }\bibfield  {title} {\enquote {\bibinfo {title} {Measurement of
  kinetic-scale current filamentation dynamics and associated magnetic fields
  in interpenetrating plasmas},}\ }\href
  {https://doi.org/10.1103/PhysRevLett.124.215001} {\bibfield  {journal}
  {\bibinfo  {journal} {Phys. Rev. Lett.}\ }\textbf {\bibinfo {volume} {124}},\
  \bibinfo {pages} {215001} (\bibinfo {year} {2020})}\BibitemShut {NoStop}%
\bibitem [{\citenamefont {Ross}\ \emph {et~al.}(2012)\citenamefont {Ross},
  \citenamefont {Glenzer}, \citenamefont {Amendt}, \citenamefont {Berger},
  \citenamefont {Divol}, \citenamefont {Kugland}, \citenamefont {Landen},
  \citenamefont {Plechaty}, \citenamefont {Remington}, \citenamefont {Ryutov},
  \citenamefont {Rozmus}, \citenamefont {Froula}, \citenamefont {Fiksel},
  \citenamefont {Sorce}, \citenamefont {Kuramitsu}, \citenamefont {Morita},
  \citenamefont {Sakawa}, \citenamefont {Takabe}, \citenamefont {Drake},
  \citenamefont {Grosskopf}, \citenamefont {Kuranz}, \citenamefont {Gregori},
  \citenamefont {Meinecke}, \citenamefont {Murphy}, \citenamefont {Koenig},
  \citenamefont {Pelka}, \citenamefont {Ravasio}, \citenamefont {Vinci},
  \citenamefont {Liang}, \citenamefont {Presura}, \citenamefont {Spitkovsky},
  \citenamefont {Miniati},\ and\ \citenamefont {Park}}]{ross12pop}%
  \BibitemOpen
  \bibfield  {author} {\bibinfo {author} {\bibfnamefont {J.~S.}\ \bibnamefont
  {Ross}}, \bibinfo {author} {\bibfnamefont {S.~H.}\ \bibnamefont {Glenzer}},
  \bibinfo {author} {\bibfnamefont {P.}~\bibnamefont {Amendt}}, \bibinfo
  {author} {\bibfnamefont {R.}~\bibnamefont {Berger}}, \bibinfo {author}
  {\bibfnamefont {L.}~\bibnamefont {Divol}}, \bibinfo {author} {\bibfnamefont
  {N.~L.}\ \bibnamefont {Kugland}}, \bibinfo {author} {\bibfnamefont {O.~L.}\
  \bibnamefont {Landen}}, \bibinfo {author} {\bibfnamefont {C.}~\bibnamefont
  {Plechaty}}, \bibinfo {author} {\bibfnamefont {B.}~\bibnamefont {Remington}},
  \bibinfo {author} {\bibfnamefont {D.}~\bibnamefont {Ryutov}}, \bibinfo
  {author} {\bibfnamefont {W.}~\bibnamefont {Rozmus}}, \bibinfo {author}
  {\bibfnamefont {D.~H.}\ \bibnamefont {Froula}}, \bibinfo {author}
  {\bibfnamefont {G.}~\bibnamefont {Fiksel}}, \bibinfo {author} {\bibfnamefont
  {C.}~\bibnamefont {Sorce}}, \bibinfo {author} {\bibfnamefont
  {Y.}~\bibnamefont {Kuramitsu}}, \bibinfo {author} {\bibfnamefont
  {T.}~\bibnamefont {Morita}}, \bibinfo {author} {\bibfnamefont
  {Y.}~\bibnamefont {Sakawa}}, \bibinfo {author} {\bibfnamefont
  {H.}~\bibnamefont {Takabe}}, \bibinfo {author} {\bibfnamefont {R.~P.}\
  \bibnamefont {Drake}}, \bibinfo {author} {\bibfnamefont {M.}~\bibnamefont
  {Grosskopf}}, \bibinfo {author} {\bibfnamefont {C.}~\bibnamefont {Kuranz}},
  \bibinfo {author} {\bibfnamefont {G.}~\bibnamefont {Gregori}}, \bibinfo
  {author} {\bibfnamefont {J.}~\bibnamefont {Meinecke}}, \bibinfo {author}
  {\bibfnamefont {C.~D.}\ \bibnamefont {Murphy}}, \bibinfo {author}
  {\bibfnamefont {M.}~\bibnamefont {Koenig}}, \bibinfo {author} {\bibfnamefont
  {A.}~\bibnamefont {Pelka}}, \bibinfo {author} {\bibfnamefont
  {A.}~\bibnamefont {Ravasio}}, \bibinfo {author} {\bibfnamefont
  {T.}~\bibnamefont {Vinci}}, \bibinfo {author} {\bibfnamefont
  {E.}~\bibnamefont {Liang}}, \bibinfo {author} {\bibfnamefont
  {R.}~\bibnamefont {Presura}}, \bibinfo {author} {\bibfnamefont
  {A.}~\bibnamefont {Spitkovsky}}, \bibinfo {author} {\bibfnamefont
  {F.}~\bibnamefont {Miniati}},\ and\ \bibinfo {author} {\bibfnamefont {H.-S.}\
  \bibnamefont {Park}},\ }\bibfield  {title} {\enquote {\bibinfo {title}
  {Characterizing counter-streaming interpenetrating plasmas relevant to
  astrophysical collisionless shocks},}\ }\href
  {https://doi.org/10.1063/1.3694124} {\bibfield  {journal} {\bibinfo
  {journal} {Physics of Plasmas}\ }\textbf {\bibinfo {volume} {19}},\ \bibinfo
  {pages} {056501} (\bibinfo {year} {2012})}\BibitemShut {NoStop}%
\bibitem [{\citenamefont {Morita}\ \emph {et~al.}(2013)\citenamefont {Morita},
  \citenamefont {Sakawa}, \citenamefont {Tomita}, \citenamefont {Ide},
  \citenamefont {Kuramitsu}, \citenamefont {Nishio}, \citenamefont {Nakayama},
  \citenamefont {Inoue}, \citenamefont {Moritaka}, \citenamefont {Ide},
  \citenamefont {Kuwada}, \citenamefont {Tsubouchi}, \citenamefont {Uchino},\
  and\ \citenamefont {Takabe}}]{morita13pop}%
  \BibitemOpen
  \bibfield  {author} {\bibinfo {author} {\bibfnamefont {T.}~\bibnamefont
  {Morita}}, \bibinfo {author} {\bibfnamefont {Y.}~\bibnamefont {Sakawa}},
  \bibinfo {author} {\bibfnamefont {K.}~\bibnamefont {Tomita}}, \bibinfo
  {author} {\bibfnamefont {T.}~\bibnamefont {Ide}}, \bibinfo {author}
  {\bibfnamefont {Y.}~\bibnamefont {Kuramitsu}}, \bibinfo {author}
  {\bibfnamefont {K.}~\bibnamefont {Nishio}}, \bibinfo {author} {\bibfnamefont
  {K.}~\bibnamefont {Nakayama}}, \bibinfo {author} {\bibfnamefont
  {K.}~\bibnamefont {Inoue}}, \bibinfo {author} {\bibfnamefont
  {T.}~\bibnamefont {Moritaka}}, \bibinfo {author} {\bibfnamefont
  {H.}~\bibnamefont {Ide}}, \bibinfo {author} {\bibfnamefont {M.}~\bibnamefont
  {Kuwada}}, \bibinfo {author} {\bibfnamefont {K.}~\bibnamefont {Tsubouchi}},
  \bibinfo {author} {\bibfnamefont {K.}~\bibnamefont {Uchino}},\ and\ \bibinfo
  {author} {\bibfnamefont {H.}~\bibnamefont {Takabe}},\ }\bibfield  {title}
  {\enquote {\bibinfo {title} {Thomson scattering measurement of a shock in
  laser-produced counter-streaming plasmas},}\ }\href
  {https://doi.org/10.1063/1.4821967} {\bibfield  {journal} {\bibinfo
  {journal} {Physics of Plasmas}\ }\textbf {\bibinfo {volume} {20}},\ \bibinfo
  {pages} {092115} (\bibinfo {year} {2013})}\BibitemShut {NoStop}%
\bibitem [{\citenamefont {Rinderknecht}\ \emph {et~al.}(2018)\citenamefont
  {Rinderknecht}, \citenamefont {Park}, \citenamefont {Ross}, \citenamefont
  {Amendt}, \citenamefont {Higginson}, \citenamefont {Wilks}, \citenamefont
  {Haberberger}, \citenamefont {Katz}, \citenamefont {Froula}, \citenamefont
  {Hoffman}, \citenamefont {Kagan}, \citenamefont {Keenan},\ and\ \citenamefont
  {Vold}}]{rinderknecht18prl}%
  \BibitemOpen
  \bibfield  {author} {\bibinfo {author} {\bibfnamefont {H.~G.}\ \bibnamefont
  {Rinderknecht}}, \bibinfo {author} {\bibfnamefont {H.-S.}\ \bibnamefont
  {Park}}, \bibinfo {author} {\bibfnamefont {J.~S.}\ \bibnamefont {Ross}},
  \bibinfo {author} {\bibfnamefont {P.~A.}\ \bibnamefont {Amendt}}, \bibinfo
  {author} {\bibfnamefont {D.~P.}\ \bibnamefont {Higginson}}, \bibinfo {author}
  {\bibfnamefont {S.~C.}\ \bibnamefont {Wilks}}, \bibinfo {author}
  {\bibfnamefont {D.}~\bibnamefont {Haberberger}}, \bibinfo {author}
  {\bibfnamefont {J.}~\bibnamefont {Katz}}, \bibinfo {author} {\bibfnamefont
  {D.~H.}\ \bibnamefont {Froula}}, \bibinfo {author} {\bibfnamefont {N.~M.}\
  \bibnamefont {Hoffman}}, \bibinfo {author} {\bibfnamefont {G.}~\bibnamefont
  {Kagan}}, \bibinfo {author} {\bibfnamefont {B.~D.}\ \bibnamefont {Keenan}},\
  and\ \bibinfo {author} {\bibfnamefont {E.~L.}\ \bibnamefont {Vold}},\
  }\bibfield  {title} {\enquote {\bibinfo {title} {Highly resolved measurements
  of a developing strong collisional plasma shock},}\ }\href
  {https://doi.org/10.1103/PhysRevLett.120.095001} {\bibfield  {journal}
  {\bibinfo  {journal} {Phys. Rev. Lett.}\ }\textbf {\bibinfo {volume} {120}},\
  \bibinfo {pages} {095001} (\bibinfo {year} {2018})}\BibitemShut {NoStop}%
\bibitem [{\citenamefont {Derouillat}\ \emph {et~al.}(2018)\citenamefont
  {Derouillat}, \citenamefont {Beck}, \citenamefont {Pérez}, \citenamefont
  {Vinci}, \citenamefont {Chiaramello}, \citenamefont {Grassi}, \citenamefont
  {Flé}, \citenamefont {Bouchard}, \citenamefont {Plotnikov}, \citenamefont
  {Aunai}, \citenamefont {Dargent}, \citenamefont {Riconda},\ and\
  \citenamefont {Grech}}]{derouillat18cpc}%
  \BibitemOpen
  \bibfield  {author} {\bibinfo {author} {\bibfnamefont {J.}~\bibnamefont
  {Derouillat}}, \bibinfo {author} {\bibfnamefont {A.}~\bibnamefont {Beck}},
  \bibinfo {author} {\bibfnamefont {F.}~\bibnamefont {Pérez}}, \bibinfo
  {author} {\bibfnamefont {T.}~\bibnamefont {Vinci}}, \bibinfo {author}
  {\bibfnamefont {M.}~\bibnamefont {Chiaramello}}, \bibinfo {author}
  {\bibfnamefont {A.}~\bibnamefont {Grassi}}, \bibinfo {author} {\bibfnamefont
  {M.}~\bibnamefont {Flé}}, \bibinfo {author} {\bibfnamefont {G.}~\bibnamefont
  {Bouchard}}, \bibinfo {author} {\bibfnamefont {I.}~\bibnamefont {Plotnikov}},
  \bibinfo {author} {\bibfnamefont {N.}~\bibnamefont {Aunai}}, \bibinfo
  {author} {\bibfnamefont {J.}~\bibnamefont {Dargent}}, \bibinfo {author}
  {\bibfnamefont {C.}~\bibnamefont {Riconda}},\ and\ \bibinfo {author}
  {\bibfnamefont {M.}~\bibnamefont {Grech}},\ }\bibfield  {title} {\enquote
  {\bibinfo {title} {Smilei : A collaborative, open-source, multi-purpose
  particle-in-cell code for plasma simulation},}\ }\href
  {https://doi.org/https://doi.org/10.1016/j.cpc.2017.09.024} {\bibfield
  {journal} {\bibinfo  {journal} {Computer Physics Communications}\ }\textbf
  {\bibinfo {volume} {222}},\ \bibinfo {pages} {351--373} (\bibinfo {year}
  {2018})}\BibitemShut {NoStop}%
\bibitem [{\citenamefont {Arber}\ \emph {et~al.}(2015)\citenamefont {Arber},
  \citenamefont {Bennett}, \citenamefont {Brady}, \citenamefont
  {Lawrence-Douglas}, \citenamefont {Ramsay}, \citenamefont {Sircombe},
  \citenamefont {Gillies}, \citenamefont {Evans}, \citenamefont {Schmitz},
  \citenamefont {Bell},\ and\ \citenamefont {Ridgers}}]{arber15ppcf}%
  \BibitemOpen
  \bibfield  {author} {\bibinfo {author} {\bibfnamefont {T.~D.}\ \bibnamefont
  {Arber}}, \bibinfo {author} {\bibfnamefont {K.}~\bibnamefont {Bennett}},
  \bibinfo {author} {\bibfnamefont {C.~S.}\ \bibnamefont {Brady}}, \bibinfo
  {author} {\bibfnamefont {A.}~\bibnamefont {Lawrence-Douglas}}, \bibinfo
  {author} {\bibfnamefont {M.~G.}\ \bibnamefont {Ramsay}}, \bibinfo {author}
  {\bibfnamefont {N.~J.}\ \bibnamefont {Sircombe}}, \bibinfo {author}
  {\bibfnamefont {P.}~\bibnamefont {Gillies}}, \bibinfo {author} {\bibfnamefont
  {R.~G.}\ \bibnamefont {Evans}}, \bibinfo {author} {\bibfnamefont
  {H.}~\bibnamefont {Schmitz}}, \bibinfo {author} {\bibfnamefont {A.~R.}\
  \bibnamefont {Bell}},\ and\ \bibinfo {author} {\bibfnamefont {C.~P.}\
  \bibnamefont {Ridgers}},\ }\bibfield  {title} {\enquote {\bibinfo {title}
  {Contemporary particle-in-cell approach to laser-plasma modelling},}\ }\href
  {https://doi.org/10.1088/0741-3335/57/11/113001} {\bibfield  {journal}
  {\bibinfo  {journal} {Plasma Physics and Controlled Fusion}\ }\textbf
  {\bibinfo {volume} {57}},\ \bibinfo {pages} {113001} (\bibinfo {year}
  {2015})}\BibitemShut {NoStop}%
\bibitem [{\citenamefont {Kuramitsu}\ \emph {et~al.}(2016)\citenamefont
  {Kuramitsu}, \citenamefont {Mizuta}, \citenamefont {Sakawa}, \citenamefont
  {Tanji}, \citenamefont {Ide}, \citenamefont {Sano}, \citenamefont {Koenig},
  \citenamefont {Ravasio}, \citenamefont {Pelka}, \citenamefont {Takabe},
  \citenamefont {Gregory}, \citenamefont {Woolsey}, \citenamefont {Moritaka},
  \citenamefont {Matsukiyo}, \citenamefont {Matsumoto},\ and\ \citenamefont
  {Ohnishi}}]{kuramitsu16apj}%
  \BibitemOpen
  \bibfield  {author} {\bibinfo {author} {\bibfnamefont {Y.}~\bibnamefont
  {Kuramitsu}}, \bibinfo {author} {\bibfnamefont {A.}~\bibnamefont {Mizuta}},
  \bibinfo {author} {\bibfnamefont {Y.}~\bibnamefont {Sakawa}}, \bibinfo
  {author} {\bibfnamefont {H.}~\bibnamefont {Tanji}}, \bibinfo {author}
  {\bibfnamefont {T.}~\bibnamefont {Ide}}, \bibinfo {author} {\bibfnamefont
  {T.}~\bibnamefont {Sano}}, \bibinfo {author} {\bibfnamefont {M.}~\bibnamefont
  {Koenig}}, \bibinfo {author} {\bibfnamefont {A.}~\bibnamefont {Ravasio}},
  \bibinfo {author} {\bibfnamefont {A.}~\bibnamefont {Pelka}}, \bibinfo
  {author} {\bibfnamefont {H.}~\bibnamefont {Takabe}}, \bibinfo {author}
  {\bibfnamefont {C.~D.}\ \bibnamefont {Gregory}}, \bibinfo {author}
  {\bibfnamefont {N.}~\bibnamefont {Woolsey}}, \bibinfo {author} {\bibfnamefont
  {T.}~\bibnamefont {Moritaka}}, \bibinfo {author} {\bibfnamefont
  {S.}~\bibnamefont {Matsukiyo}}, \bibinfo {author} {\bibfnamefont
  {Y.}~\bibnamefont {Matsumoto}},\ and\ \bibinfo {author} {\bibfnamefont
  {N.}~\bibnamefont {Ohnishi}},\ }\bibfield  {title} {\enquote {\bibinfo
  {title} {Time evolution of {K}elvin{\textendash}{H}elmholtz vortices
  associated with collisionless shocks in laser-produced plasmas},}\ }\href
  {https://doi.org/10.3847/0004-637x/828/2/93} {\bibfield  {journal} {\bibinfo
  {journal} {The Astrophysical Journal}\ }\textbf {\bibinfo {volume} {828}},\
  \bibinfo {pages} {93} (\bibinfo {year} {2016})}\BibitemShut {NoStop}%
\bibitem [{\citenamefont {Sakai}\ \emph {et~al.}(2022)\citenamefont {Sakai},
  \citenamefont {Moritaka}, \citenamefont {Morita}, \citenamefont {Tomita},
  \citenamefont {Minami}, \citenamefont {Nishimoto}, \citenamefont {Egashira},
  \citenamefont {Ota}, \citenamefont {Sakawa}, \citenamefont {Ozaki},
  \citenamefont {Kodama}, \citenamefont {Kojima}, \citenamefont {Takezaki},
  \citenamefont {Yamazaki}, \citenamefont {Tanaka}, \citenamefont {Aihara},
  \citenamefont {Koenig}, \citenamefont {Albertazzi}, \citenamefont {Mabey},
  \citenamefont {Woolsey}, \citenamefont {Matsukiyo}, \citenamefont {Takabe},
  \citenamefont {Hoshino},\ and\ \citenamefont {Kuramitsu}}]{sakai22srep}%
  \BibitemOpen
  \bibfield  {author} {\bibinfo {author} {\bibfnamefont {K.}~\bibnamefont
  {Sakai}}, \bibinfo {author} {\bibfnamefont {T.}~\bibnamefont {Moritaka}},
  \bibinfo {author} {\bibfnamefont {T.}~\bibnamefont {Morita}}, \bibinfo
  {author} {\bibfnamefont {K.}~\bibnamefont {Tomita}}, \bibinfo {author}
  {\bibfnamefont {T.}~\bibnamefont {Minami}}, \bibinfo {author} {\bibfnamefont
  {T.}~\bibnamefont {Nishimoto}}, \bibinfo {author} {\bibfnamefont
  {S.}~\bibnamefont {Egashira}}, \bibinfo {author} {\bibfnamefont
  {M.}~\bibnamefont {Ota}}, \bibinfo {author} {\bibfnamefont {Y.}~\bibnamefont
  {Sakawa}}, \bibinfo {author} {\bibfnamefont {N.}~\bibnamefont {Ozaki}},
  \bibinfo {author} {\bibfnamefont {R.}~\bibnamefont {Kodama}}, \bibinfo
  {author} {\bibfnamefont {T.}~\bibnamefont {Kojima}}, \bibinfo {author}
  {\bibfnamefont {T.}~\bibnamefont {Takezaki}}, \bibinfo {author}
  {\bibfnamefont {R.}~\bibnamefont {Yamazaki}}, \bibinfo {author}
  {\bibfnamefont {S.~J.}\ \bibnamefont {Tanaka}}, \bibinfo {author}
  {\bibfnamefont {K.}~\bibnamefont {Aihara}}, \bibinfo {author} {\bibfnamefont
  {M.}~\bibnamefont {Koenig}}, \bibinfo {author} {\bibfnamefont
  {B.}~\bibnamefont {Albertazzi}}, \bibinfo {author} {\bibfnamefont
  {P.}~\bibnamefont {Mabey}}, \bibinfo {author} {\bibfnamefont
  {N.}~\bibnamefont {Woolsey}}, \bibinfo {author} {\bibfnamefont
  {S.}~\bibnamefont {Matsukiyo}}, \bibinfo {author} {\bibfnamefont
  {H.}~\bibnamefont {Takabe}}, \bibinfo {author} {\bibfnamefont
  {M.}~\bibnamefont {Hoshino}},\ and\ \bibinfo {author} {\bibfnamefont
  {Y.}~\bibnamefont {Kuramitsu}},\ }\bibfield  {title} {\enquote {\bibinfo
  {title} {Direct observations of pure electron outflow in magnetic
  reconnection},}\ }\href {https://doi.org/10.1038/s41598-022-14582-3}
  {\bibfield  {journal} {\bibinfo  {journal} {Scientific Reports}\ }\textbf
  {\bibinfo {volume} {12}},\ \bibinfo {pages} {10921} (\bibinfo {year}
  {2022})}\BibitemShut {NoStop}%
\bibitem [{\citenamefont {Kato}\ and\ \citenamefont
  {Takabe}(2010)}]{kato10pop}%
  \BibitemOpen
  \bibfield  {author} {\bibinfo {author} {\bibfnamefont {T.~N.}\ \bibnamefont
  {Kato}}\ and\ \bibinfo {author} {\bibfnamefont {H.}~\bibnamefont {Takabe}},\
  }\bibfield  {title} {\enquote {\bibinfo {title} {Electrostatic and
  electromagnetic instabilities associated with electrostatic shocks:
  Two-dimensional particle-in-cell simulation},}\ }\href
  {https://doi.org/10.1063/1.3372138} {\bibfield  {journal} {\bibinfo
  {journal} {Physics of Plasmas}\ }\textbf {\bibinfo {volume} {17}},\ \bibinfo
  {pages} {032114} (\bibinfo {year} {2010})}\BibitemShut {NoStop}%
\bibitem [{\citenamefont {Ohira}\ and\ \citenamefont
  {Takahara}(2008)}]{ohira08apj}%
  \BibitemOpen
  \bibfield  {author} {\bibinfo {author} {\bibfnamefont {Y.}~\bibnamefont
  {Ohira}}\ and\ \bibinfo {author} {\bibfnamefont {F.}~\bibnamefont
  {Takahara}},\ }\bibfield  {title} {\enquote {\bibinfo {title} {Oblique ion
  two-stream instability in the foot region of a collisionless shock},}\ }\href
  {https://doi.org/10.1086/592182} {\bibfield  {journal} {\bibinfo  {journal}
  {The Astrophysical Journal}\ }\textbf {\bibinfo {volume} {688}},\ \bibinfo
  {pages} {320--326} (\bibinfo {year} {2008})}\BibitemShut {NoStop}%
\end{thebibliography}%

\end{document}